\documentclass[twocolumn]{aastex63}

\usepackage{graphicx}
\usepackage{amssymb,amsmath,amsfonts}
\usepackage{xcolor}

\usepackage{enumerate}
\usepackage{comment}

\newcommand{\result}[1]{#1}

\newcommand{\externalresult}[1]{#1}

\newcommand{\FourteenMassCut}{\result{\ensuremath{5.0\,M_\odot}}}
\newcommand{\FourteenQCut}{\result{\ensuremath{0.28}}}

\newcommand{\FourteenPowerlawPeakPvalCoarseGrained}{\result{\ensuremath{\leq 0.056 \% }}} % 0.0005577464909294338
\newcommand{\FourteenPowerlawPeakPvalNminusOne}{\result{\ensuremath{\leq 0.024 \%}}} % 0.00023855667117966828
\newcommand{\TwelveQCut}{\result{\ensuremath{0.8}}}

\newcommand{\TwelvePowerlawPeakPvalCoarseGrained}{\result{\ensuremath{22\%}}} % 2.215346e-01
\newcommand{\TwelvePowerlawPeakPvalNminusOne}{\result{\ensuremath{19\%}}} % 1.859432e-01

\newcommand{\TwentyOneMassCut}{\result{\ensuremath{50\,M_\odot}}}

\newcommand{\TwentyOnePowerlawPeakPvalCoarseGrained}{\result{\ensuremath{20\%}}} % 1.983131e-01
\newcommand{\TwentyOnePowerlawPeakPvalNminusOne}{\result{\ensuremath{15\%}}} % 1.490348e-01

\newcommand{\TwentyOneTruncatedPvalCoarseGrained}{\result{\ensuremath{11\%}}} % 1.108522e-01
\newcommand{\TwentyOneTruncatedPvalNminusOne}{\result{\ensuremath{6\%}}} % 6.105181e-02

\begin{document}
%----------------------------------------------------------------------------------------

\title{
Probing Extremal Gravitational-Wave Events with Coarse-Grained Likelihoods
}

\author{Reed Essick}
\email{reed.essick@gmail.com}
\affiliation{Perimeter Institute for Theoretical Physics, 31 Caroline Street North, Waterloo, Ontario, Canada, N2L 2Y5}

\author{Amanda Farah}
\affiliation{Department of Physics, University of Chicago, Chicago, IL 60637, USA}

\author{Shanika Galaudage}
\affiliation{School of Physics and Astronomy, Monash University, Clayton VIC 3800, Australia}
\affiliation{OzGrav: The ARC Centre of Excellence for Gravitational Wave Discovery, Clayton VIC 3800, Australia}

\author{Colm Talbot}
\affiliation{LIGO Laboratory, California Institute of Technology, Pasadena, CA 91125, USA}

\author{Maya Fishbach}
\altaffiliation{NASA Hubble Fellowship Program Einstein Postdoctoral Fellow}
\affiliation{Center for Interdisciplinary Exploration and Research in Astrophysics (CIERA) and Department of Physics and Astronomy,
Northwestern University, 1800 Sherman Ave, Evanston, IL 60201, USA}

\author{Eric Thrane}
\affiliation{School of Physics and Astronomy, Monash University, Clayton VIC 3800, Australia}
\affiliation{OzGrav: The ARC Centre of Excellence for Gravitational Wave Discovery, Clayton VIC 3800, Australia}

\author{Daniel E. Holz}
\affiliation{Department of Physics, University of Chicago, Chicago, IL 60637, USA}
\affiliation{Department of Astronomy and Astrophysics, Enrico Fermi Institute, and Kavli Institute for Cosmological Physics,\\University of Chicago, Chicago, IL 60637, USA}

\date{\today}

\begin{abstract}
    As catalogs of gravitational-wave transients grow, new records are set for the most extreme systems observed to date.
    The most massive observed black holes probe the physics of pair instability supernovae while providing clues about the environments in which binary black hole systems are assembled.
    The least massive black holes, meanwhile, allow us to investigate the purported neutron star-black hole mass gap, and binaries with unusually asymmetric mass ratios or large spins inform our understanding of binary and stellar evolution.
    Existing outlier tests generally implement leave-one-out analyses, but these do not account for the fact that the event being left out was by definition an extreme member of the population.
    This results in a bias in the evaluation of outliers.
    We correct for this bias by introducing a coarse-graining framework to investigate whether these extremal events are true outliers or whether they are consistent with the rest of the observed population.
    Our method enables us to study extremal events while testing for population model misspecification.
    We show that this ameliorates biases present in the leave-one-out analyses commonly used within the gravitational-wave community.
    Applying our method to results from the second LIGO--Virgo transient catalog, we find qualitative agreement with the conclusions of~\citet{O3aRatesAndPop}.
    GW190814 is an outlier because of its small secondary mass.
    We find that neither GW190412 nor GW190521 are outliers.
\end{abstract}

%----------------------------------------------------------------------------------------

\section{Introduction}
\label{sec:introduction}
As catalogs of gravitational-wave (GW) sources observed with the Advanced LIGO~\citep{LIGO} and Virgo~\citep{Virgo} interferometers continue to grow, our knowledge of the population of compact objects is continually refined.
The most recent update from the LIGO--Virgo--KAGRA (LVK) collaborations~\citep[GWTC-2;][]{GWTC-2} brings to light several interesting features within the distributions of masses and spins of compact objects in coalescing binary systems~\citep{O3aRatesAndPop}.
In particular, GWTC-2 set new records for the largest black hole mass~\citep[GW190521,][]{GW190521}, smallest black hole mass~\citep[GW190814,][]{GW190814}\footnote{It is possible that the secondary object in GW190814 is actually an unusually massive neutron star~\citep{Essick2020}.}, and most asymmetric mass ratios~\citep[GW190814 and GW190412,][]{GW190412, GW190814}.
Of immediate interest is whether these objects are merely the most extreme events observed from a single population: the most extreme examples in a catalog become more extreme as the size of the catalog grows~\citep{Fishbach2020}.
Alternatively, these events may be inconsistent with the population inferred from the rest of the detected events, and thus are \emph{true outliers}.
The observation of such outliers could suggest the first example from an as-of-yet unmodeled or entirely new (sub)population.
However, it may simply indicate that the current phenomenological population models are simply a poor description of nature.

The interpretation of the most extreme GW events has significant astrophysical implications.
The most massive binary black hole (BBH) events probe the pair-instability supernova (PISN) mass gap~\citep{2017ApJ...851L..25F,2018ApJ...856..173T}, a theoretically-proposed dearth of black holes between $\sim50$--$120\,M_\odot$~\citep{2002ApJ...567..532H}.
Observing BBH systems near the pair-instability gap can inform our knowledge of nuclear reaction rates~\citep{Farmer2020}, beyond standard model physics~\citep{Croon2020,Baxter2021}, or the boundaries of mass gaps in general~\citep{Fishbach2020Nov, Edelman2021, Nitz2021, Ezquiaga2021} and applications thereof~\citep[e.g.,][]{Farr2019}.
At the other extreme, BBH with small component masses can inform our knowledge of supernova physics~\citep[e.g.,][]{Fryer2001,2012ApJ...757...91B,Zevin2020}. 

Several authors have posited that the most massive black holes of GWTC-2, which appear to sit in the PISN mass gap, form a separate population from the ``main population'' observed to date, invoking formation scenarios such as hierarchical mergers~\citep[e.g.,][]{GW190521Properties,Kimball2021,Tagawa2021,2021NatAs.tmp..136G} or primordial black holes~\citep[e.g.,][]{2021arXiv210503349F,2021PhRvL.126e1101D}.
Indeed, in order to explain the wide range of observed BBH properties, many authors have argued that multiple formation channels are active and that the BBH population consists of multiple subpopulations~\citep{O3aRatesAndPop,Ng2020, Zevin2021}.
More than anything else, the variety of interpretations of the GWTC-2 events highlight the excitement within the GW community as new discoveries are routinely made whenever more data is recorded.
However, they also emphasize the need for thoughtful consideration of the methods employed to assess whether individual events are consistent with existing population models.

Motivated by the ``leave-one-out'' consistency checks in \citet{2019ApJ...882L..24A}, \citet{Fishbach2020}, \citet{O3aRatesAndPop}, and elsewhere, we introduce a general procedure to investigate the effect of individual events on inferred populations.
Specifically, we derive a ``coarse-grained'' analysis that retains, in a controlled way, only a subset of the total information available about some detected events.
We are then able to determine whether the population inferred from this ``coarse-grained'' data is consistent with the population inferred from the original data.

In particular, previous approaches compared populations inferred with all $N$ events in a catalog to populations inferred with only $N-1$ events, arguing that if these were similar then the event that was ``left out'' is consistent with the rest of the population, as its inclusion does not significantly change the inferred population.
If the contrary were true, the event would be considered a true outlier.
However, such tests do not account for the way in which the excluded event was selected.
It was left out specifically because it was extremal in some dimension, and this selection may artificially inflate the apparent significance of differences in inferred populations, particularly in the presence of sharp boundaries within the population model. 
Our approach improves upon this by self-consistently accounting for how the selected events are chosen in a controlled way while clearly defining the subset of the available information that is retained.
We show that our approach naturally reproduces previous techniques when a random event is excluded from the analysis.
It is precisely because the excluded events are typically not chosen at random that previous techniques can introduce biases.

There is already a healthy literature proposing leave-one-out analyses as tests for model misspecification within Bayesian inference.
For example,~\citet{Vehtari2017} constructs a cross-validation likelihood from analyses that leave out each event in a catalog one at a time.
This approach attempts to limit the possible biases associated with leaving out only extremal events, similar to the motivation for our coarse-grained approach. 
Other authors have considered tempering the likelihood in order to combat model misspecification, which is at times referred to as coarsening~\citep{Miller2015}.
In this approach, the likelihood is raised to the power of an inverse temperature $\beta = 1/T \in (0, 1]$, thereby artificially inflating statistical uncertainties.
Several procedures exist to select $\beta$~\citep[e.g.,][and references therein]{Miller2015, Thomas2019}, all of which attempt to optimize the balance between systematic error from model misspecification and additional statistical error from tempered likelihoods.
Our coarse-grained likelihoods are similar in spirit, but differ in that we eschew the use of \textit{ad hoc} tempering in favor of marginalizing over the data and parameters from an individual event subject to a precisely specified (but looser) constraint on the event's parameters.
That is, we specify the size and placement of the ``coarse grain'' used to approximate the event's likelihood.

Using this coarse-grained likelihood, we revisit several astrophysically interesting events from GWTC-2 that were discussed in detail in~\citet{O3aRatesAndPop}.
To wit, we demonstrate that
\begin{itemize}
    \item GW190814, with a secondary mass of $m_2 \sim 2.6\,M_\odot$ and mass ratio $q=m_2/m_1 \sim 0.1$, is an outlier in $m_2$ (too small to be consistent with the main BBH population) but is not an outlier in mass ratio.
    \item GW190412 is not an outlier, and its mass ratio~\citep[$q \sim 0.28$;][]{GW190412} is in the tail of the main BBH population.
    \item GW190521 is not an outlier with respect to the main BBH population under the preferred phenomenological mass models considered in~\citet{O3aRatesAndPop}. It is only marginally inconsistent under the simplest mass model considered, which is disfavored for other reasons as well.
\end{itemize}
While we believe the quantitative details of our analysis improve upon previous methods, all of our resulting astrophysical conclusions are in agreement with~\citet{O3aRatesAndPop}.

The remainder of this paper is organized as follows.
We derive our coarse-grained leave-one-out formalism from first principles in Section~\ref{sec:methods} and explore a toy-model in Section~\ref{sec:toy models} to gain intuition.
Readers only interested in our astrophysical conclusions can skip directly to Section~\ref{sec:results}, where we apply the method to public LVK data. We conclude in Section~\ref{sec:discussion}.

%----------------------------------------------------------------------------------------

\section{Coarse-Grained Leave-One-Out Likelihoods}
\label{sec:methods}

We derive a procedure for coarse-graining our inference of the population in Sec.~\ref{sec:derivation}, while in Sec.~\ref{sec:choice of region} we discuss the implications of different possible procedures to choose the amount of information retained in the coarse-grained inference.
We describe how to quantify these consistency tests with a single statistic in Sec.~\ref{sec:methods pvalues}.

%---------------------------------------------

\subsection{Derivation of Coarse-Grained Inference}
\label{sec:derivation}

To begin, let us assume we have $N$ events that are each described by a set of $m$ parameters $\theta \in \mathcal{R}^m$.
We further assume these events belong to the same astrophysical population described by the hyperparameters $\Lambda_0$:
\begin{equation}
    \theta_i \sim p(\theta|\Lambda_0) \quad \forall \quad i.
\end{equation}
If we write our model for the differential Poisson number density of signals in the universe as $d\mathcal{N}/d\theta = R p(\theta|\Lambda)$, our joint distribution over the data $\{D_i\}$, single-event parameters $\{\theta_i\}$, hyperparameters, and the rate $R$ is then
\begin{widetext}
\begin{equation}\label{eq:joint}
    p(\{D_i\};\{\theta_i\};\Lambda, R) = p(\Lambda)p(R)R^N e^{-R\mathcal{E}(\Lambda)} \prod\limits_{i=1}^N p(D_i|\theta_i)p(\theta_i|\Lambda)\Theta(\rho(D_i) \geq \rho_\mathrm{thr}) ,
\end{equation}
\end{widetext}
where $\Theta$ is the Heaviside function, $\rho(D)$ is the detection statistic used to select events for the catalog with threshold $\rho_\mathrm{thr}$, and 
\begin{align}\label{eq:selection function}
    \mathcal{E}(\Lambda) & = \int d\theta\, p(\theta|\Lambda) \int dD\, p(D|\theta) \Theta(\rho(D)\geq\rho_\mathrm{thr}) \nonumber \\
    & = \int d\theta\, p(\theta|\Lambda) P(\mathrm{det}|\theta) \nonumber \\
    & = P(\mathrm{det}|\Lambda)
\end{align}
is the expected fraction of events that are detectable within a population.
Note that Eq.~\ref{eq:joint} is neither a likelihood function nor a posterior distribution, but instead is a joint distribution for both the data and the parameters.
Furthermore, if we assume a uniform-in-log prior for the rate ($p(R) \sim 1/R$) and marginalize, we obtain
\begin{multline}\label{eq:full joint}
    p(\{D_i\};\{\theta_i\};\Lambda) = \\
        p(\Lambda) \prod\limits_{i=1}^N\left[\frac{p(D_i|\theta_i)p(\theta_i|\Lambda)\Theta(\rho(D_i) \geq \rho_\mathrm{thr})}{\mathcal{E}(\Lambda)}\right]
\end{multline}
We note that the $\Theta$ in each factor within the product does not impact the inference as it is guaranteed to be one; the data is axiomatically detectable for detected events.
However, these factors are important in what follows.

If we wish to construct a posterior for only  $\Lambda$, we marginalize over $\{\theta_i\}$ and condition on the observed data to obtain
\begin{equation}\label{eq:full post}
    p(\Lambda|\{D_i\}) \propto p(\Lambda)\prod\limits_{i=1}^N\left[ \frac{\int d\theta_i\, p(D_i|\theta_i)p(\theta_i|\Lambda)}{\mathcal{E}(\Lambda)} \right] .
\end{equation}
All these expressions have become commonplace within the GW community, and we refer the reader to the many reviews in the literature for more details~\cite[e.g.,][]{Mandel2016, Thrane2019, Vitale2020}.

We are particularly interested in the effect that subsets of events have on the inferred population, and whether those effects can be used to determine if particular events are outliers.
However, if we simply remove potential outliers, we may bias the inferred population by preferentially removing the most extreme events.
This, in turn, may artificially inflate the significance of any changes in the inferred population.
As such, we introduce a coarse-graining procedure to replace traditional ``leave-one-out" analyses.
This procedure retains only a limited amount of information about a particular event.
In this way, the analysis can naturally account for how the event-of-interest was selected (e.g., as the most extreme event in some dimension) while still discarding as much information about that event as possible.

To wit, let us assume that the $j^\mathrm{th}$ event is almost certainly within some region of parameter space $\mathcal{S}$ so that
\begin{equation}\label{eq:minimum S}
    \int d\theta_j p(\theta_j | D_j, \Lambda) \Theta(\theta_j \in \mathcal{S}) \geq 1 - \epsilon \quad \forall \quad \{\Lambda : p(\Lambda) \neq 0\}.
\end{equation}
In other words, for any population described by $\Lambda$ with nonvanishing support in the hyperprior $p(\Lambda)$, the inferred posterior on the event parameters $\theta_j$ is contained within the region $\mathcal{S}$ with high ($1-\epsilon$) credibility.
We are typically interested in the case $\epsilon \ll 1$, which is to say we are confident that event $j$ is contained within ${\cal S}$.
There is a uniquely defined smallest $\mathcal{S}$ that satisfies this requirement, which in general depends on the choice of $\epsilon$ (smaller $\epsilon$ require larger $\mathcal{S}$).
Additionally, we only assert that $\mathcal{S}$ is at least as large as the minimal region; it may be larger.
Choosing a larger region corresponds to retaining less information about the $j^\mathrm{th}$ event.

At times, it is possible to identify individual events that are clearly separated from the rest of the population, in which case it may be straightforward to identify an appropriate choice for $\mathcal{S}$.
An example could be defining $\mathcal{S}$ as the region with masses smaller than the minimum mass observed in a set.
However, more complicated boundaries are possible, as hypersurfaces in multi-dimensional spaces could divide events along a nontrivial slice that depends on multiple parameters simultaneously.
For example, we may define $\mathcal{S}$ to be a region with a secondary mass or mass ratio smaller than the corresponding minima observed in a set (see Sec.~\ref{sec:GW190814}).
We discuss several procedures to choose $\mathcal{S}$ in Sec.~\ref{sec:choice of region}.

Of course, an extremal event is not necessarily inconsistent with the population determined by the other $N-1$ events; some event is always the most extreme of any set.
Indeed, the $j^\mathrm{th}$ event may still be drawn from the same $p(\theta|\Lambda_0)$ and simply be the most extreme example in the catalog.
We therefore test the null-hypothesis that all $N$ events are drawn from the same distribution by comparing the inferred population when we include all $N$ events to the population inferred from $N-1$ events while accounting for the knowledge that $\theta_j \in \mathcal{S}$.
If the population inferred from the coarse-grained analysis is inconsistent with the full dataset, we reject the null hypothesis that the $j^\mathrm{th}$ event is drawn from the same distribution as the other $N-1$ events.
With this method, we can be assured that we do not overestimate the significance of differences in the inferred populations because we include knowledge that the $j^\mathrm{th}$ event was selected in a particular way.

We now consider how to self-consistently limit the information about the $j^\mathrm{th}$ event within our inference.
Because $\theta_j \in \mathcal{S}$ almost surely ($\epsilon \ll 1$), we can add a term to Eq.~\ref{eq:full joint} without affecting the overall inference for $\Lambda$:
\begin{widetext}
\begin{equation}
\label{eq:coarse-grained full}
    p(\{D_i\};\{\theta_i\};\Lambda) \simeq p(\Lambda) \left( \frac{p(D_j|\theta_j)p(\theta_j|\Lambda)\Theta(\rho(D_j)\geq \rho_\mathrm{thr})\Theta(\theta_j \in \mathcal{S})}{\mathcal{E}(\Lambda)} \right) \prod\limits_{i\neq j}^{N-1} \left[\frac{p(D_i|\theta_i)p(\theta_i|\Lambda)\Theta(\rho(D_i)\geq\rho_\mathrm{thr})}{\mathcal{E}(\Lambda)}\right]
\end{equation}
\end{widetext}
where we have included an additional factor of $\Theta(\theta_j \in \mathcal{S})$ since this is one almost everywhere there is posterior support for $\theta_j$ (to within the precision specified in Eq.~\ref{eq:minimum S}).
This is completely analogous to the way in which terms representing the detectability of data, i.e., $\Theta(\rho(D) \geq \rho_\mathrm{thr})$, are always one for the detected events.
We have also replaced ``='' with ``$\simeq$'' to denote that strict equality only holds in the limit $\epsilon \rightarrow 0$.
However, in what follows, we assume that $\epsilon$ is small enough to be negligible and retain the ``='' in what follows.

Furthermore, the only information we wish to retain about the $j^\mathrm{th}$ event is that it almost certainly is within $\mathcal{S}$, and therefore we marginalize over both $D_j$ and $\theta_j$ to effectively forget everything else about the event.
This yields
\begin{widetext}
\begin{align}
    \int d D_j d\theta_j\, p(D_j|\theta_j) p(\theta_j|\Lambda) \Theta(\rho(D_j) \geq \rho_\mathrm{thr}) \Theta(\theta_j \in \mathcal{S})
        & = \int d\theta_j\, \Theta(\theta_j \in \mathcal{S}) p(\theta_j|\Lambda) \int d D_j p(D_j|\theta_j) \Theta(\rho(D_j) \geq \rho_\mathrm{thr}) \nonumber \\
        & = \int d\theta_j\, \Theta(\theta_j \in \mathcal{S}) p(\theta_j|\Lambda) P(\mathrm{det}|\theta_j) \nonumber \\
        & = P(\mathrm{det}, \theta_j \in \mathcal{S}|\Lambda) ,
\end{align}
\end{widetext}
which is just the probability that the event was detected and had parameters within $\mathcal{S}$ given the underlying population model.
We additionally note that this term only appears in Eq.~\ref{eq:coarse-grained full} within a ratio, and the divisor of that ratio is $\mathcal{E}(\Lambda) = P(\mathrm{det}|\Lambda)$.
Therefore the coarse-graining procedure yields an overall factor of
\begin{equation}\label{eq:cofactor}
    \frac{P(\mathrm{det},\theta_j\in\mathcal{S}|\Lambda)}{P(\mathrm{det}|\Lambda)} = P(\theta_j \in \mathcal{S}|\mathrm{det}, \Lambda) .
\end{equation}
This has the appealing interpretation that the only information retained about the $j^\mathrm{th}$ event is the probability that it belongs to a particular part of parameter space (our ``coarse grain'') given that it was detected and came from a particular population.\footnote{Note that Eq.~\ref{eq:cofactor} refers to the probability that a detected event from a population came from $\mathcal{S}$, whereas Eq.~\ref{eq:minimum S} refers the the probability that a specific event (the $j^\mathrm{th}$ observed event, which produced data $D_j$) came from $\mathcal{S}$.}
We discuss implications and implicit assumptions made by the choice of $\mathcal{S}$ in more detail in Sec.~\ref{sec:choice of region}.
Briefly, we note that most ``agnostic'' procedures for defining $\mathcal{S}$ will not depend on $D_j$ or $\theta_j$, and so we explicitly write $\mathcal{S}=\mathcal{S}(\{\theta_{i \neq j}\})$ below.

Putting everything together, we marginalize the coarse-grained likelihood over $\{\theta_{i \neq j}\}$ and condition on $\{D_{i \neq j}\}$ to obtain a coarse-grained posterior for $\Lambda$.
\begin{widetext}
\begin{equation}\label{eq:coarse-grained post}
    p(\Lambda|\{D_{i \neq j}\}, \rho(D_j) \geq \rho_\mathrm{thr}; \theta_j \in \mathcal{S}(\{\theta_{i \neq j}\})) \propto p(\Lambda) \int \left( \left[\prod\limits_{i \neq j}^{N-1} d\theta_i\right] \left[\prod\limits_{k \neq j}^{N-1} \frac{p(D_k|\theta_k) p(\theta_k|\Lambda)}{\mathcal{E}(\Lambda)}\right] P(\theta_j \in \mathcal{S}(\{\theta_{i \neq j}\}) |\mathrm{det}, \Lambda) \right)
\end{equation}
\end{widetext}
We note that the marginalization over $\{\theta_{i \neq j}\}$ does not factor as nicely as it does in Eq.~\ref{eq:full post} because $\mathcal{S}(\{\theta_{i \neq j}\})$ may depend on all the events (except the $j^\mathrm{th}$ event) and is included within the integral.
Methods to calculate Eq.~\ref{eq:coarse-grained post} efficiently when one already has access to samples from $p(\Lambda|\{D_{i \neq j}\})$ or $p(\Lambda|\{D_i\})$ are discussed in Appendix~\ref{sec:reweighing}.

%---------------------------------------------

\subsection{Implications from the Choice of $\mathcal{S}$}
\label{sec:choice of region}

Sec.~\ref{sec:derivation} presents the coarse-grained inference for $\Lambda$ based on knowledge that $\theta_j \in \mathcal{S}$.
This holds for arbitrary $\mathcal{S}$ as long as Eq.~\ref{eq:minimum S} is satisfied; the choice of $\mathcal{S}$ is up to the analyst.
We now consider the implications of different choices for $\mathcal{S}$ and what assumptions they represent.

To begin, if $\theta_j$ is well determined by $D_j$ and the minimum allowable $\mathcal{S}$ is small (particularly in comparison to the extent of the prior $p(\theta|\Lambda)$), the coarse-graining procedure can still retain most, if not all, the relevant information about the $j^\mathrm{th}$ event.
In this way, defining $\mathcal{S}$ in relation to the minimum allowable $\mathcal{S}$ implicitly assumes some knowledge of $D_j$, which is undesirable for a leave-one-out analysis.
To avoid this, we instead recommend choosing $\mathcal{S}$ based on only \textit{a priori} theoretical predictions of astrophysical interest or the data from the $N-1$ other events.
In general, this implies that one should write $\mathcal{S} = \mathcal{S}(\{\theta_{i \neq j}\})$ to explicitly show that it does not depend on either the data or parameters from the $j^\mathrm{th}$ event, as we do within Eq.~\ref{eq:coarse-grained post}.

One possible data-driven procedure for choosing $\mathcal{S}$ is particularly appealing in the special case where the parameters of the $j^\mathrm{th}$ event are cleanly separated from the rest of the events.
If the $j^\mathrm{th}$ event is the most extreme event in some dimension $x$, we can simply define $\mathcal{S}(\{x_{i\neq j}\}): x \geq \max_{i \neq j}\,\{x_i\}$ as the region where $x$ is larger than the second-largest detected event.\footnote{One could equivalently consider the smallest event.}
This choice is natural in the sense that it only depends on $\{D_{i \neq j}\}$ and is as generous as possible subject to the knowledge that the $j^\mathrm{th}$ event is extremal; it does not retain any information about how much larger $x_j$ is than $\max_{i \neq j}\,\{x_i\}$.

Such a choice allows an analyst to pose questions such as, ``how big should the largest individual event in a catalog of $N$ events be given the observation of $N-1$ events and the knowledge that one was larger?''
Comparing the predicted largest event to what was actually observed naturally defines a metric that can be used as a quantitative consistency check (see Sec.~\ref{sec:methods pvalues}).
Indeed, the choice of $\mathcal{S}$ precisely specifies the information used when computing $p$-values for the null hypothesis that all $N$ events in a catalog were drawn from the same population.
Nonetheless, it is important to remember that this procedure is not unique, just as the definition of a null hypothesis is not unique, and other choices of $\mathcal{S}$ may be able to more naturally answer other questions.

In this vein, one might also consider choosing $\mathcal{S}$ based on prior theoretical motivations, so long as the selected region is compatible with Eq.~\ref{eq:minimum S}.
This can provide a perfectly natural way to define alternative tests of the null hypothesis and is particularly relevant when the events are not cleanly separated.
For example, even before observing any data, we may identify the region $m_1 > 50\,M_\odot$ as interesting from the standpoint of PISN theory, and scrutinize any events that fall in this region as potential outliers.

We note that the inability to define $\mathcal{S}$ based on the next-most extreme event only arises when the $j^\mathrm{th}$ event is not cleanly separated from the other events, and therefore it is not unambiguously the most extreme event.
One may take that ambiguity itself as evidence that the event cannot be an outlier, and therefore argue that our machinery may not be needed in such situations.
Nonetheless, we can still construct a perfectly self-consistent coarse-grained inference even when the minimal $\mathcal{S}$ surrounding the $j^\mathrm{th}$ event's parameters overlaps significantly with the inferred parameters of the other $N-1$ events.
The coarse-grained event need not actually be extremal.

As a limiting case, we note that choosing $\mathcal{S}$ that spans the entire parameter space ($\mathcal{S} \rightarrow \mathcal{R}^m$) implies that $P(\theta_j \in S | \mathrm{det}, \Lambda) \rightarrow 1 \ \forall \ \Lambda$.
That is, if we know nothing about the parameters of the $j^\mathrm{th}$ event, then the coarse-grained inference is equivalent to neglecting that event altogether and performing the ``standard'' inference for $\Lambda$ with the other $N-1$ events.\footnote{The corresponding inference of the rate would still correctly account for the fact that we detected $N$ events in total, which is not necessarily true of other leave-one-out analyses.}
This is intuitive in that, if we picked an event to coarse-grain at random without first considering its parameters, then the smallest allowable $\mathcal{S}$ that was certain to contain the event would be the entire (detectable) parameter space.
At the same time, excluding an event at random should not introduce any biases within the inference for $\Lambda$, and one expects to perform the standard inference with only $N-1$ events.
We see, then, that performing a standard inference with $N-1$ events after excluding an extremal event, \emph{without} incorporating the knowledge that the excluded event was extremal, is not a self-consistent procedure because it does not correctly reflect the data selection procedure.
This inconsistency leads to biases, which we demonstrate with a toy model in Sec.~\ref{sec:toy models}.

%---------------------------------------------

\subsection{Quantifying Inconsistencies with Inferred Populations}
\label{sec:methods pvalues}

In order to test the null hypothesis that the event in question is consistent with the $N-1$ other events, we follow a similar procedure to~\citet{Fishbach2020} and construct a $p$-value statistic.
We marginalize over both the uncertainty in the population hyperparameters, inferred from the coarse-grained inference, and the measurement uncertainty in the parameters of the $N$ observations.
For each hyperposterior sample $\Lambda$ from the coarse-grained inference, we
\begin{enumerate}[i)]
    \item draw $N$ synthetic detected events from the corresponding population described by those hyperparameters,
    \item reweigh the $N$ observed events' posteriors for $\{\theta_{i}\}$ to what would be obtained under the population prior described by that $\Lambda$. For the $N-1$ events that were not excluded, this recovers their marginal posterior for $\{\theta_{i\neq j}\}$ under the joint distribution of Eq.~\ref{eq:coarse-grained full}, while for the excluded event we obtain the population-informed posterior for $\theta_j$,
    \item draw one $\theta_\mathrm{obs}$ sample for each observed event, and 
    \item compare the most extreme $\theta_\mathrm{syn}$ out of the $N$ synthetic events to the most extreme $\theta_\mathrm{obs}$.
\end{enumerate}
Repeating this procedure for many hyperposterior samples $\Lambda$, the $p$-value is obtained as the fraction of synthetic catalogs that produce an event that is at least as extreme as the most extreme observed event.
Although such statistics are not necessarily uniformly distributed even under the null hypothesis (see, e.g.,~\citet{10.1214/18-BA1124}), if the resulting marginalized $p$-value is small, we reject the null hypothesis and conclude that the excluded event is inconsistent with the main population.

\citet{Fishbach2020} estimated $p$-values by scattering the maximum likelihood estimates by synthetic noise realizations and then comparing the maximum likelihood from the observed event to the synthetic distribution under the null hypothesis.
Our approach differs in that we do not scatter the maximum likelihood estimate of our synthetic detections, but instead use the true values of the detected events.
However, we still marginalize over the actual uncertainty in the observed events' true parameters stemming from detector noise.
Both procedures, then, account for measurement uncertainty but answer slightly different questions.

%----------------------------------------------------------------------------------------

\section{Toy Model}
\label{sec:toy models}

\begin{figure*}
    \centering
    \includegraphics[width=\textwidth, clip=True, trim=0.0cm 0.25cm 0.0cm 0.5cm]{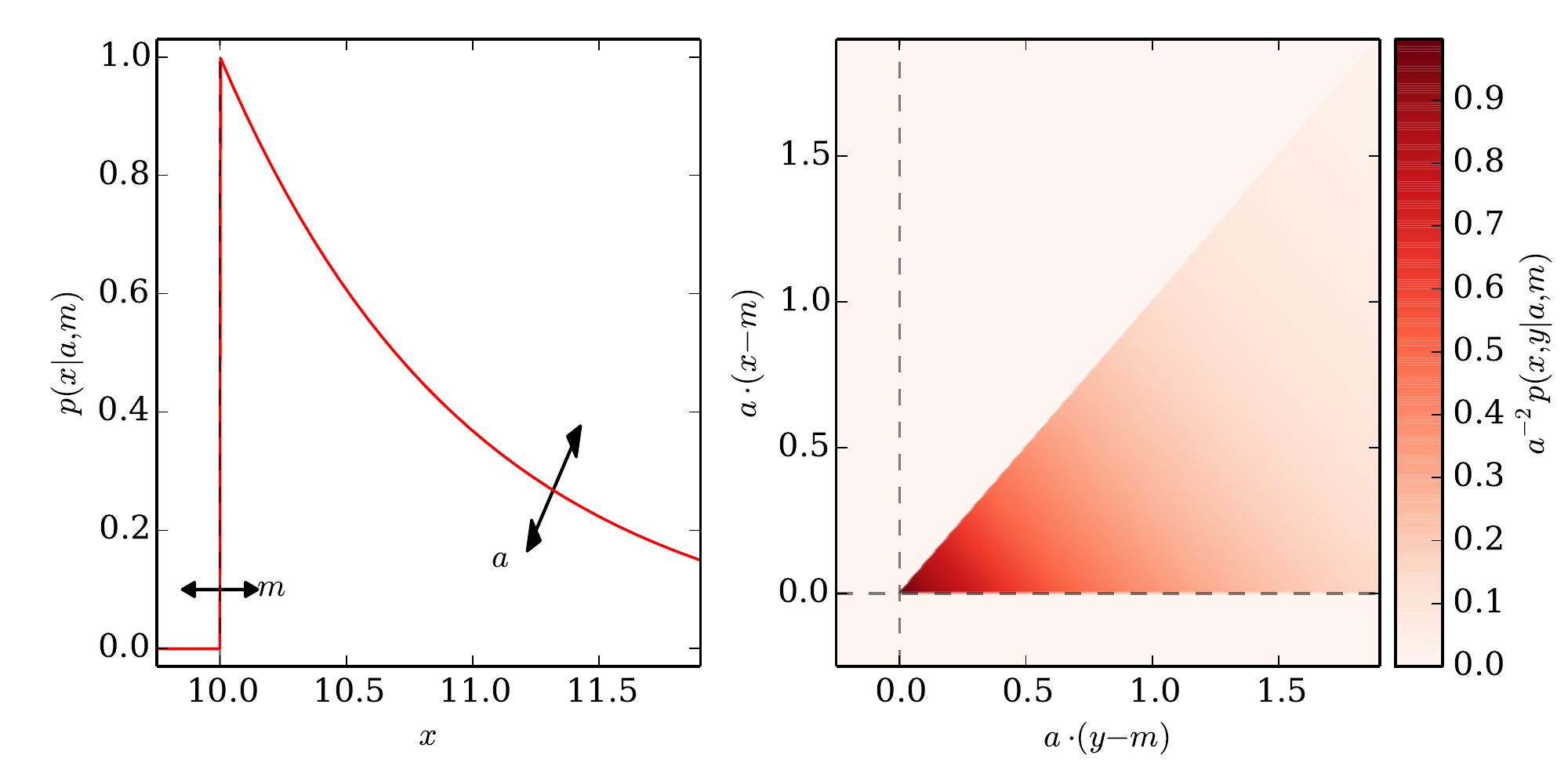}
    \caption{
        Toy population model with a sharp lower limit.
        We show (\emph{left}) the distribution of single parameters (Eq.~\ref{eq:toy model indiv_pop}), annotated to show how the hyperparameters affect the shape of the distribution, and (\emph{right}) the joint distribution (Eq.~\ref{eq:toy model pop}) from which individual events are drawn to form our synthetic catalogs.
    }
    \label{fig:toy model distribution}
\end{figure*}

We first investigate a simple toy model to gain an intuition for how our coarse-graining formalism impacts population inferences.
Specifically, we investigate the impact of defining $\mathcal{S}$ based on multiple single-event parameters in the presence of sharp edges within a population model (e.g., mass gaps or spin cutoffs).
Of particular interest is the impact on our uncertainty in the inferred location of the sharp edges~\citep[c.f.,][]{Farr2019, Ezquiaga2021}.

Fig.~\ref{fig:toy model distribution} shows our population model.
We assume individual events are described by two real parameters
\begin{align}
    x_i, y_i
        & \sim p(x, y|a, m) \nonumber \\
        & = p(x|a, m) p(y|a, m) \Theta(m \leq x \leq y)
\end{align}
where
\begin{equation}\label{eq:toy model indiv_pop}
    p(x|a, m) = a e^{-a(x-m)} \Theta(m\leq x)
\end{equation}
In this model, objects are independently drawn from the same distribution, which has a sharp cut-off at $x=m$, and are then randomly paired subject to an arbitrary labeling scheme ($x \leq y$).
This implies
\begin{equation}\label{eq:toy model pop}
    p(x, y|a, m) = 2 a^2 e^{ -a(x+y-2m) } \Theta(m\leq x \leq y)
\end{equation}
In this model, $m$ controls the smallest allowed value within the population and $a$ controls the spread in values; larger $a$ imply faster exponential decay and more tightly clustered values.
Furthermore, we assume that all events are observable ($\mathcal{E}(a,m) = 1\ \forall\ a,m$) and have vanishingly small observational uncertainties ($p(x,y|D_i) \sim \delta(x-x_i)\delta(y-y_i)$) for simplicity.
This also implies that there is no ambiguity in which event is the most extreme in any dimension.
The full hyperposterior is then
\begin{align}\label{eq:toy model hyperposterior}
    p(a, m|\{x_i,y_i\})
        & \propto p(a, m) \prod\limits_i^N p(x_i, y_i|a, m) \nonumber \\
        & \propto p(a, m) (2 a^2)^N e^{-a\sum_i^N(x_i+y_i - 2m)} \nonumber \\
        & \quad\quad\quad\quad \times \Theta(m \leq \min\limits_i\{x_i\})
\end{align}
with hyperprior $p(a,m)$.
For concreteness, we assume
\begin{gather}
    p(a) = \frac{1}{A\Gamma(\alpha+1)}\left(\frac{a}{A}\right)^\alpha e^{-a/A} \\
    p(m) = \frac{1}{M\Gamma(\mu+1)}\left(\frac{m}{M}\right)^\mu e^{-m/M}
\end{gather}
which render the hyperposterior analytically tractable.
Figs.~\ref{fig:toy model x hyperposteriors} and~\ref{fig:toy model q hyperposteriors} consider the limits $\alpha, \mu \rightarrow 0$ and $A, M \rightarrow \infty$ to obtain uninformative hyperpriors.

We also consider the coarse-grained hyperposterior, which takes the form
\begin{multline}\label{eq:toy model coarse-grained post}
    p(a, m|\{x_{i\neq j}, y_{i\neq j}\}, (x_j, y_j) \in \mathcal{S}) \propto \\ p(a,m|\{x_i,y_i\})\frac{P(x,y\in \mathcal{S}|a,m)}{p(x_j,y_j|a,m)}
\end{multline}
where
\begin{equation}
    P(x,y\in\mathcal{S}|a,m) \equiv \int\limits_\mathcal{S} dx dy\, 2a^2 e^{-a(x+y-2m)} \Theta(m\leq x \leq y)
\end{equation}
Because there is no ambiguity in which event is the most extreme, we choose $\mathcal{S}$ based on the parameters of the other $N-1$ events.
We consider two special cases: (Sec.~\ref{sec:toy model excluding smallest x}) excluding the event with the smallest $x$ and (Sec.~\ref{sec:toy model excluding smallest q}) excluding the event with the smallest $q \equiv x/y$.

%------------------------

\subsection{Excluding the Smallest $x$}
\label{sec:toy model excluding smallest x}

If we exclude the event with the smallest $x$, we obtain $\mathcal{S}: x \leq \min\limits_{i \neq j}\{x_i\} \equiv \mathcal{M}$ and
\begin{align}\label{eq:toy model x hyperposterior}
    P(x,y\in\mathcal{S}|a,m)
        & = P(x \leq \mathcal{M}|a,m) \nonumber \\
        & = \int\limits_m^\mathcal{M}dx \int\limits_x^\infty dy\, 2a^2 e^{-a(x+y-2m)} \nonumber \\
        & = \left(1 - e^{-2a(\mathcal{M} - m)}\right) \Theta(m \leq \mathcal{M})
\end{align}
From this, we can immediately write down the full hyperposterior with Eq.~\ref{eq:toy model coarse-grained post}.

\begin{figure*}
    \begin{minipage}{0.49\textwidth}
        \begin{center}
            {\Large Null Hypothesis is True} \\
            \includegraphics[width=1.0\textwidth]{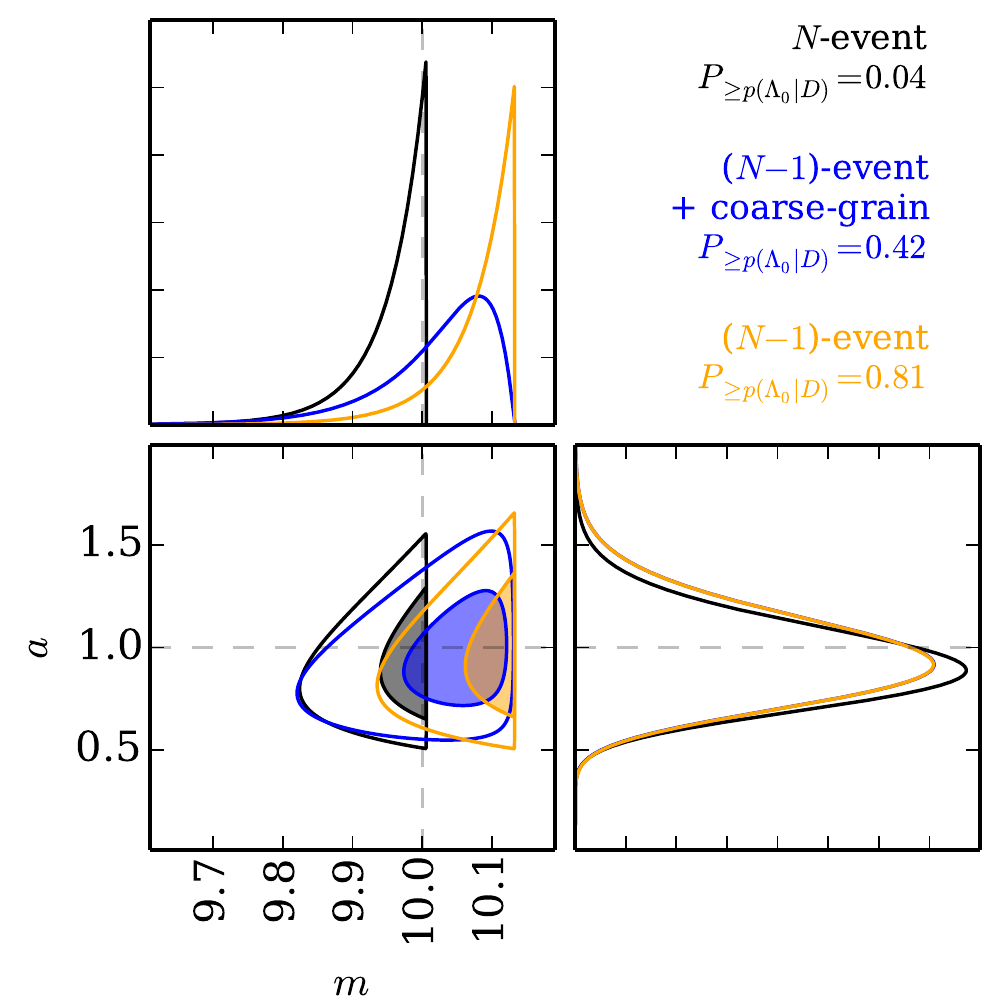} \\
            \includegraphics[width=1.0\textwidth]{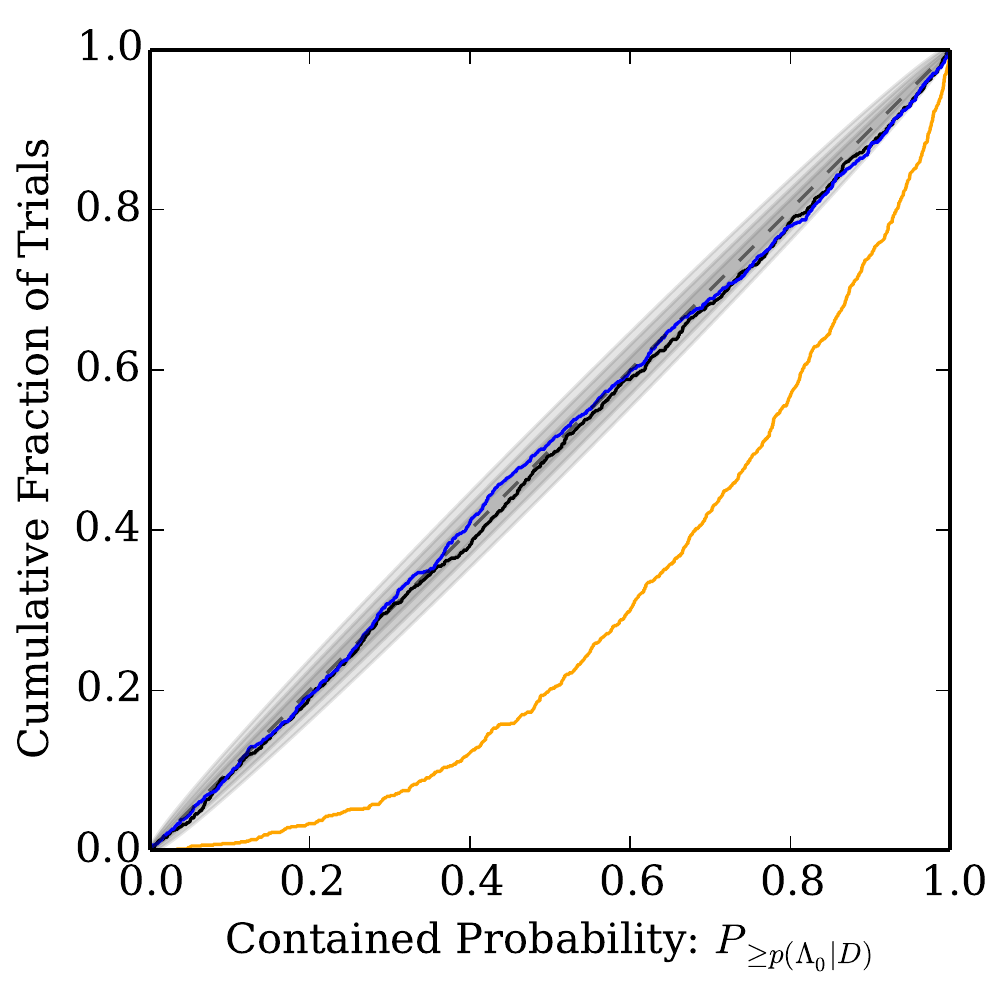}
        \end{center}
    \end{minipage}
    \begin{minipage}{0.49\textwidth}
        \begin{center}
            {\Large Null Hypothesis is False} \\
            \includegraphics[width=1.0\textwidth]{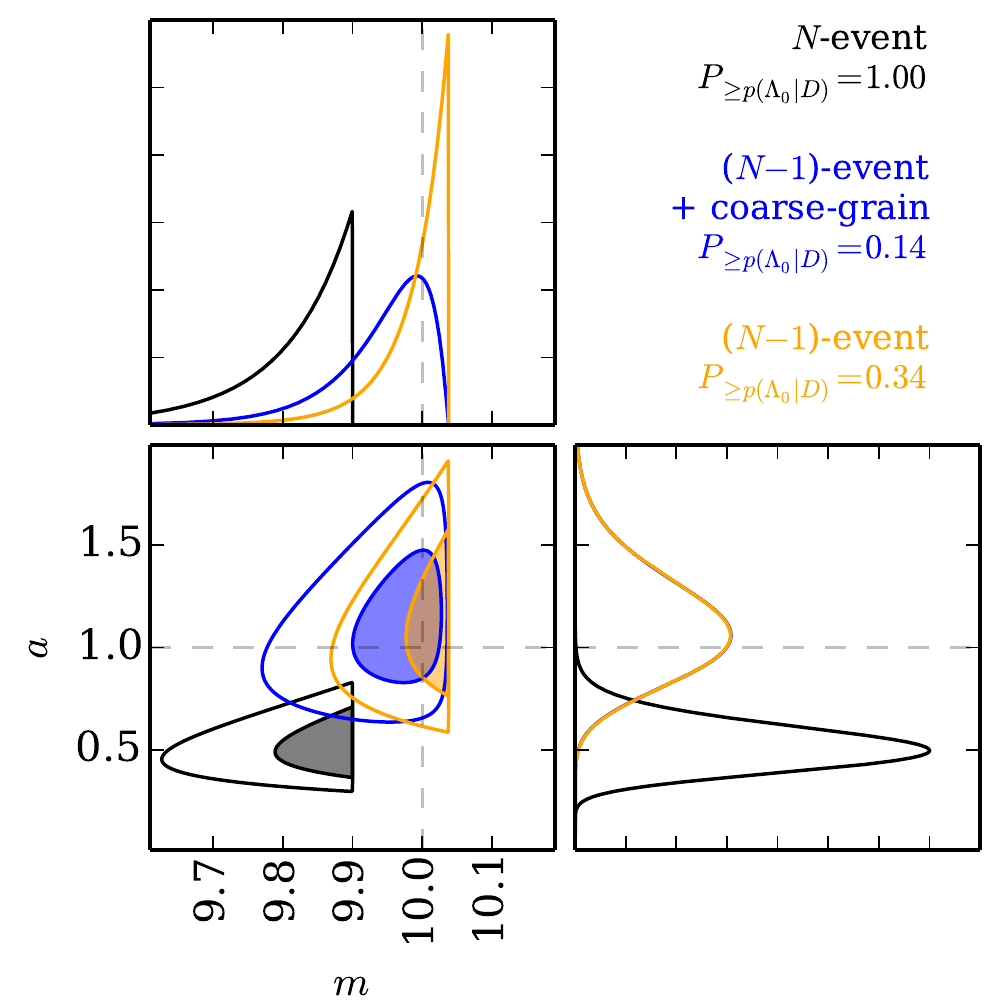} \\
            \includegraphics[width=1.0\textwidth]{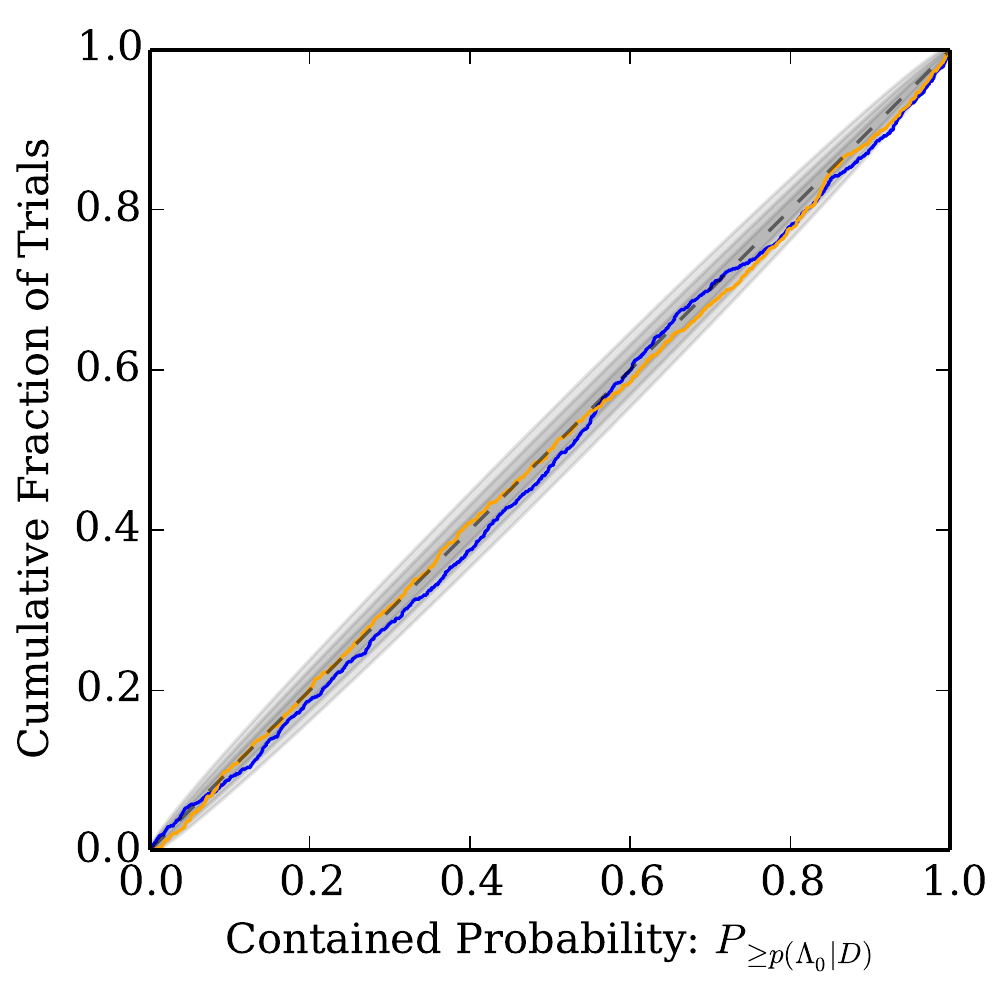}
        \end{center}
    \end{minipage}
    \caption{
        Toy Model in which we exclude the event with the smallest $x$ from catalogs of 10 events when (\emph{left}) all events are drawn from the same distribution and (\emph{right}) $N-1$ events are drawn from the same distribution and there is an additional true outlier at $(x, y) = (9.9, 30)$.
        (\emph{top row}) Example realizations of synthetic catalogs.
        Dashed lines correspond to the true hyperparameters.
        We see that the ($N-1$)-event coarse-grained inference ameliorates systematic biases in both the $N$-event and ($N-1$)-event inferences when their underlying assumptions are incorrect.
        Note that the marginal ($N-1$)-event hyperposteriors for $a$ are almost identical.
        (\emph{bottom row}) Cumulative histograms of the posterior probability integrated over the region with $p(\Lambda|\{D\}) \geq p(\Lambda_0|\{D\})$.
        Diagonal lines indicate proper coverage, and shaded regions approximate expected 1, 2, and 3-$\sigma$ deviations from the counting uncertainty with the finite number of trials performed.
        We note that the ($N-1$)-event inference introduces large systematic biases (true hyperparameters are preferentially found in the tails of the hyperposterior) when the null hypothesis is true, while the $N$-event inference may not ever contain the true hyperparameters when the null hypothesis is false (the $N$-event curve is absent from the bottom-right panel because the contained probability is always 1).
        However, in both cases, the ($N-1$)-event inference with coarse-graining provides reliable posteriors, although we do expect them to be biased to some extent when the null hypothesis is false.
    }
    \label{fig:toy model x hyperposteriors}
\end{figure*}

In our simple model, excluding the event with the smallest $x$ primarily impacts our knowledge of $m$, the minimum allowed value within the population.
Fig.~\ref{fig:toy model x hyperposteriors} shows two examples with synthetic data, one in which all simulated events are drawn from the same population (our null hypothesis) and one in which a single event is drawn from a different population centered at $x \ll m$ (a true outlier) while the other $N-1$ events are drawn from the original population.
For each example, we show the inferred hyperposterior using all $N$ events (Eq.~\ref{eq:toy model hyperposterior}), the coarse-grained hyperposterior (Eq.~\ref{eq:toy model coarse-grained post} and~\ref{eq:toy model x hyperposterior}), and the inferred posterior from Eq.~\ref{eq:toy model hyperposterior} when we use only $N-1$ events and do \emph{not} account for the fact that we excluded the event with the smallest $x$.

If the null hypothesis is true, we see that the coarse-grained hyperposterior agrees well with the hyperposterior that uses all \result{$N=10$} events.
The coarse-graining procedure correctly accounts for the additional probability associated with detecting an event with small $x$.
Conversely, the hyperposterior using $N-1$ events without the coarse-graining correction is biased towards higher $m$.
Indeed, this bias, could lead to the erroneous identification of the smallest event as inconsistent with the main population.

In addition to individual realizations of synthetic catalogs, Fig.~\ref{fig:toy model x hyperposteriors} also shows the cumulative distributions of the total probability from the region assigned a posterior probability $p(\Lambda|\{D\}) \geq p(\Lambda_0|\{D\})$, the posterior at the true hyperparameters.
Correct coverage corresponds to diagonal lines, and the shaded grey regions demonstrate the size of expected 1-, 2-, and 3-sigma fluctuations from the finite number of trials performed.
When the null hypothesis is true, we see that the coarse-grained inference is unbiased; it agrees well with the full $N$-event hyperposterior and has correct coverage.
Conversely, the ($N-1$)-event hyperposterior that does not include the coarse-graining correction does not have correct coverage; the true population parameters are systematically assigned posterior probabilities that are too low.

When the null hypothesis is false and the smallest event was not drawn from the same population as the other $N-1$ events, we see markedly different behavior.
Here, the full $N$-event hyperposterior is biased to significantly lower $m$ while both of the ($N-1$)-event inferences are much less affected.
The ($N-1$)-event inference that does not include the coarse-graining correction is unbiased and has correct coverage in this case.
It correctly excludes the extremal event from the inference of the main population.
The coarse-grained ($N-1$)-event hyperposterior is biased when the null hypothesis is incorrect, as it incorrectly assumes the $j^\mathrm{th}$ event is drawn from the same population as the other $N-1$ events, but it is much less biased than the full $N$-event result.
Indeed, it appears to have nearly correct coverage.

This suggests the following rule of thumb:
If the null hypothesis is correct, the full $N$-event hyperposterior should be very similar to the ($N-1$)-event coarse-grained hyperposterior.
However, if the null hypothesis is incorrect, the coarse-grained hyperposterior is likely to be more similar to the ($N-1$)-event hyperposterior that does not contain the coarse-graining correction.
However, this may be violated in practice (see Sec.~\ref{sec:GW190521}) and we suggest decisions be based on quantitative assessments like those proposed in Sec.~\ref{sec:methods pvalues}.
Furthermore, in both cases we note that one should not use the coarse-grained hyperposterior as the ``final inferred population.''
Even though it consistently provides a reasonable estimate of the uncertainty in the population, it is only a useful diagnostic tool to determine which of the other hyperposteriors we believe.
We also suggest estimates of $p$-values be performed with the coarse-grained hyperposterior (see Sec.~\ref{sec:pvalues}).

%------------------------

\subsection{Excluding the Smallest $q\equiv x/y$}
\label{sec:toy model excluding smallest q}

We additionally investigate the impact of discarding the event with the smallest $q$, defining $\mathcal{S}: q \leq \min\limits_{i \neq j}\{q_{i}\} \equiv \mathcal{Q}$.
With the coordinate change $q=x/y$ and $r=x+y$, we obtain
\begin{widetext}
\begin{align}
    P(x,y\in\mathcal{S}:q\leq\mathcal{Q}|a,m) & = 2 a^2 e^{2ma} \int\limits_0^\mathcal{Q} dq\, (1+q)^{-2} \int\limits_{m(1+q)/q} dr\, r e^{-ar} \nonumber \\
        & = 2\left(\frac{\mathcal{Q}}{1+\mathcal{Q}}\right) e^{ m a (\mathcal{Q}-1)/\mathcal{Q}}
\end{align}
\end{widetext}
because $r=y(1+q) = x(1+q)/q \geq m(1+q)/q$.

\begin{figure*}
    \begin{minipage}{0.49\textwidth}
        \begin{center}
            {\Large Null Hypothesis is True} \\
            \includegraphics[width=1.0\textwidth]{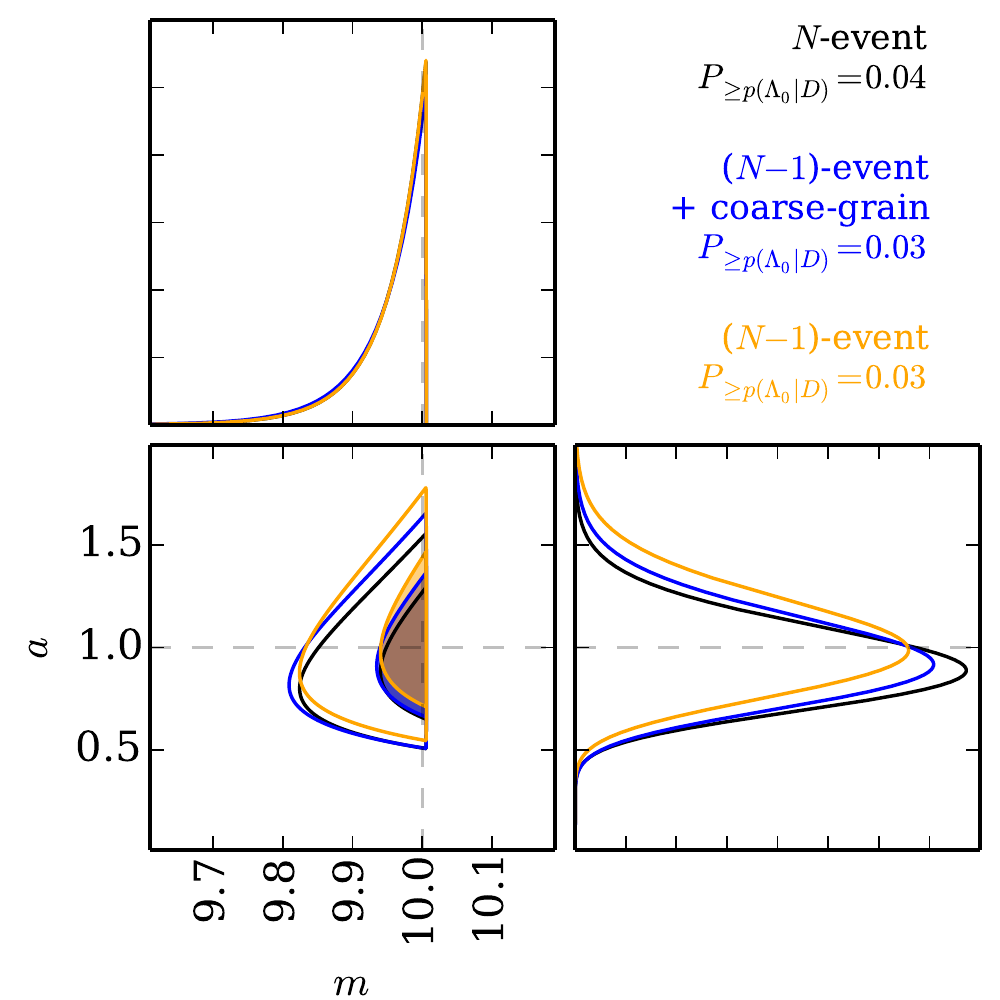} \\
            \includegraphics[width=1.0\textwidth]{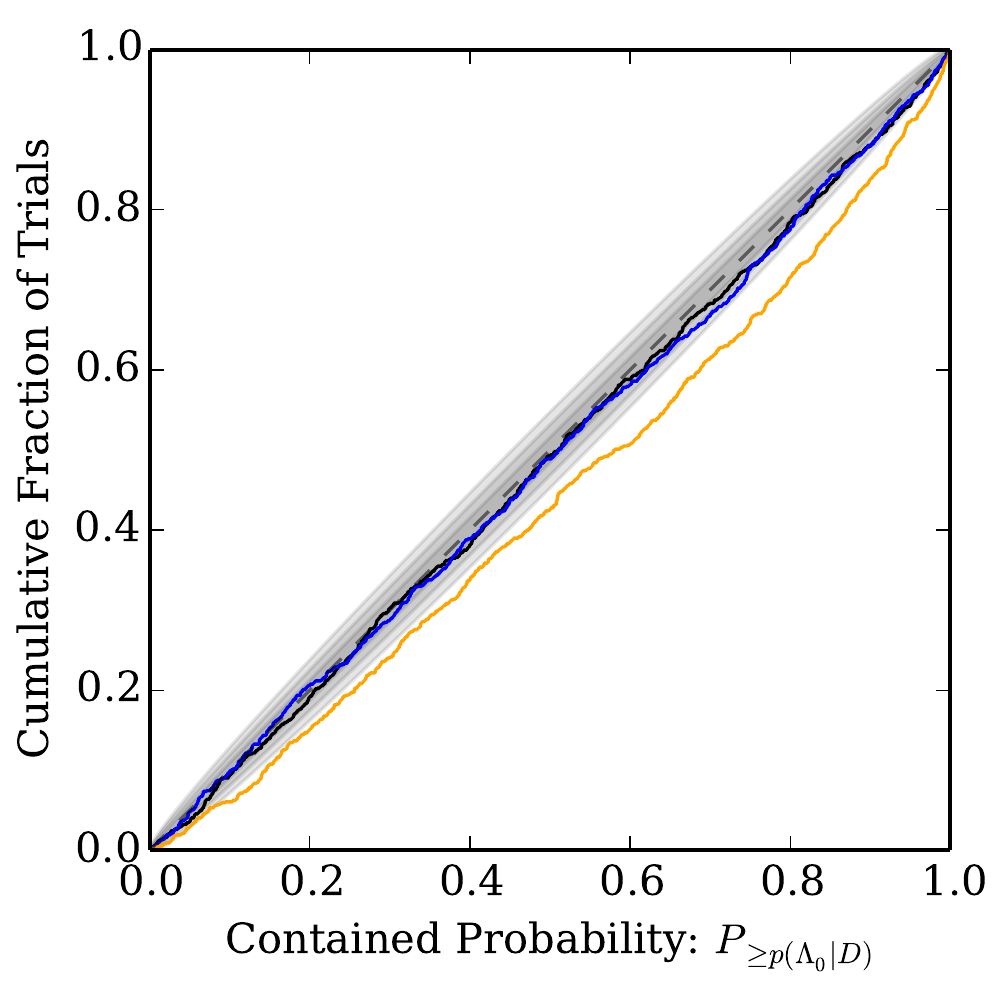}
        \end{center}
    \end{minipage}
    \begin{minipage}{0.49\textwidth}
        \begin{center}
            {\Large Null Hypothesis is False} \\
            \includegraphics[width=1.0\textwidth]{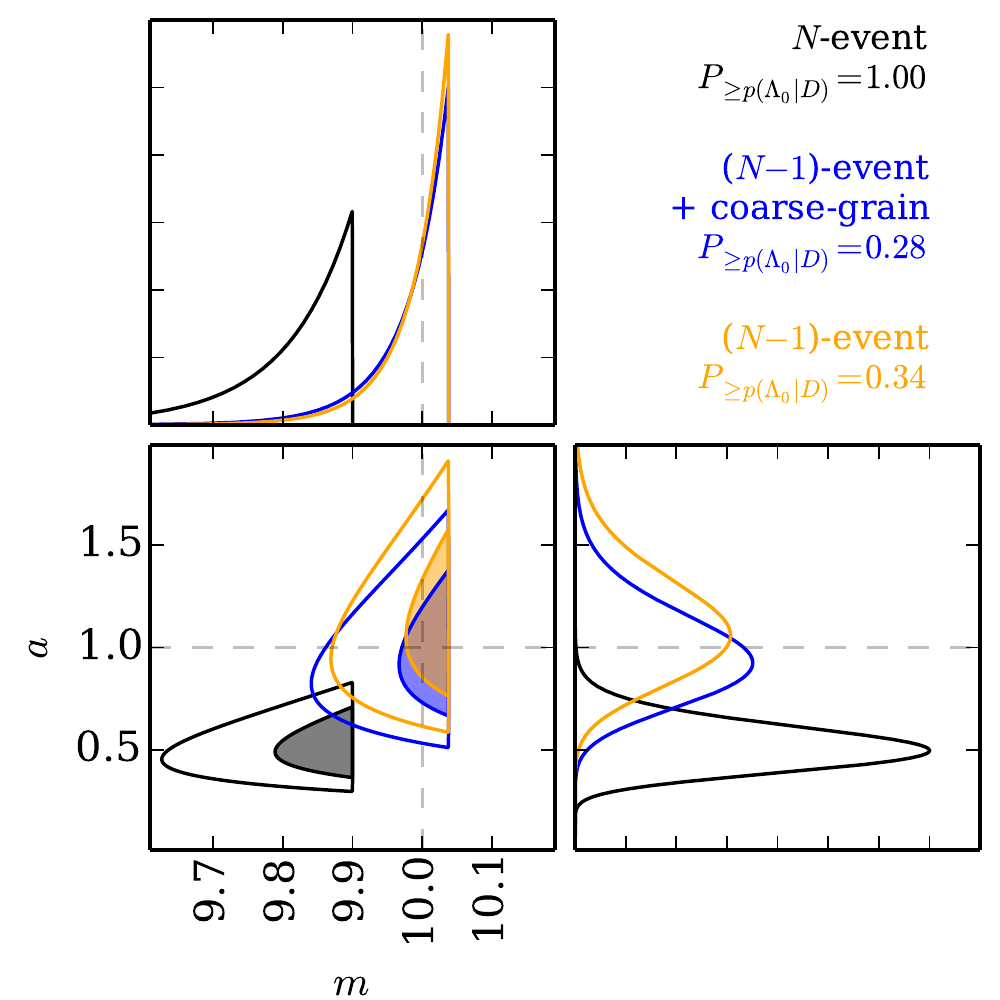} \\
            \includegraphics[width=1.0\textwidth]{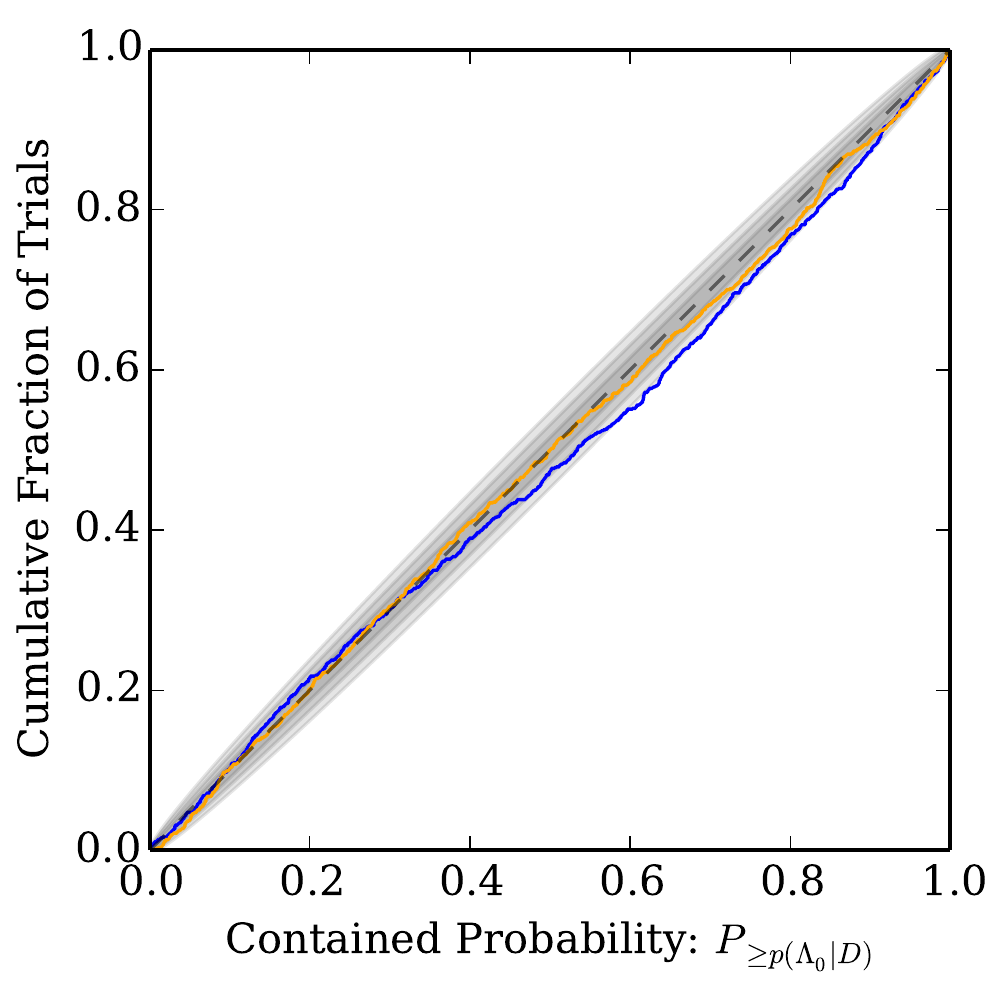}
        \end{center}
    \end{minipage}
    \caption{
        Analogous to Fig.~\ref{fig:toy model x hyperposteriors}, except simulations exclude the event with the smallest $q\equiv x/y$.
        We note that biases from incorrect assumptions appear to be smaller when we exclude events with the smallest $q$ compared to when we exclude events with the smallest $x$, which we attribute to the lack of a sharp feature in the population distribution over $q$.
        Nonetheless, we consistently observe incorrect coverage for the ($N-1$)-event hyperposterior when the null hypothesis is true at $\gtrsim 3$-$\sigma$.
    }
    \label{fig:toy model q hyperposteriors}
\end{figure*}

Fig.~\ref{fig:toy model q hyperposteriors} summarizes our conclusions.
In general, our inference of $a$ is more affected than $m$ when excluding the event with the smallest $q$.
This is because $a$ more closely controls the range of values supported in the population (larger $a$ imply faster exponential decay and more concentrated samples).
If the events are restricted to a narrow range, then they are more likely to have $q \sim 1$.
For this reason, the ($N-1$)-event hyperposterior that does not include the coarse-graining correction is biased to larger $a$ (larger $q$) when the null hypothesis is true.
Generally, the coverage is better in all cases, which we attribute to the lack of sharp features for $q$ in the population model.
As in Sec.~\ref{sec:toy model excluding smallest x}, we find that the coarse-grained hyperposterior is generally more robust against the presence (or absence) of true outliers than either of the other options.

%---------------------

\subsection{Testing the Null Hypothesis with Coarse-Grained $p$-values}
\label{sec:pvalues}

Beyond the intuition developed in Sec.~\ref{sec:toy model excluding smallest x} and~\ref{sec:toy model excluding smallest q}, we wish to quantify the probability that the full set of $N$ events are drawn from the same distribution (our null hypothesis).
Because the coarse-grained hyperposterior provides a reasonable estimate of the true underlying population regardless of whether the null hypothesis is correct, we compute $p$-values which assume individual extremal events are consistent with the population inferred within the coarse-grained inference following the procedure detailed in Sec.~\ref{sec:methods pvalues}.
Indeed, a primary motivation for our coarse-grained inference is to avoid accidentally biasing such $p$-values to higher significance by excluding extremal events, or to artificially lower their significance by computing $p$-values with the full N-event population analysis.

\begin{figure*}
    \begin{minipage}{0.49\textwidth}
        \begin{center}
            \includegraphics[width=1.0\textwidth]{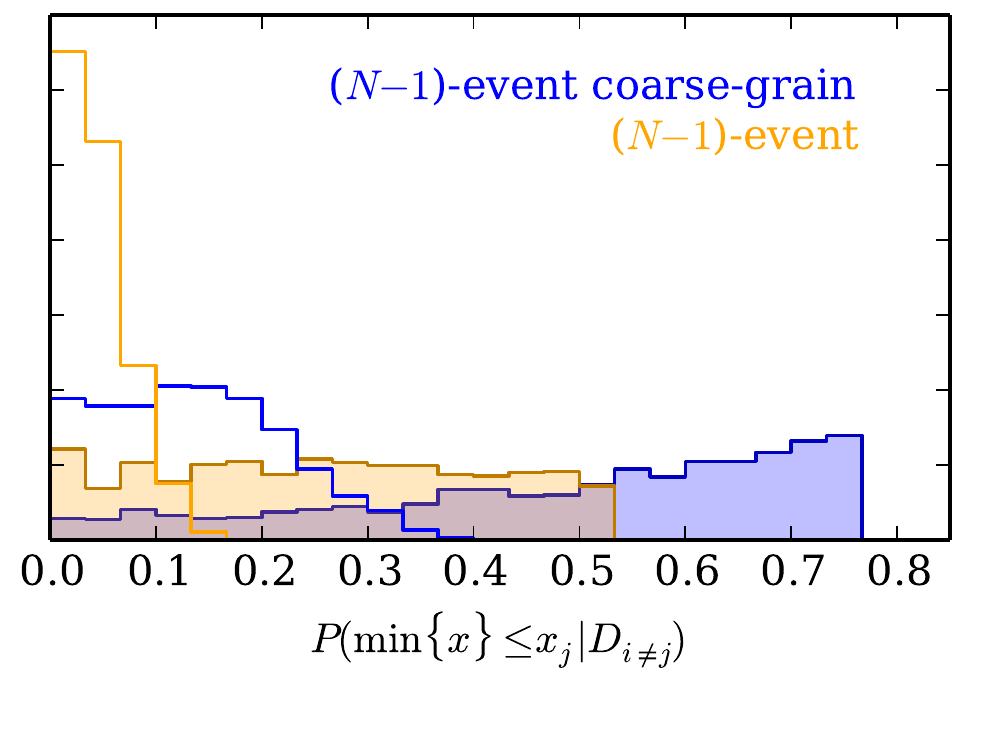}
        \end{center}
    \end{minipage}
    \begin{minipage}{0.49\textwidth}
        \begin{center}
            \includegraphics[width=1.0\textwidth]{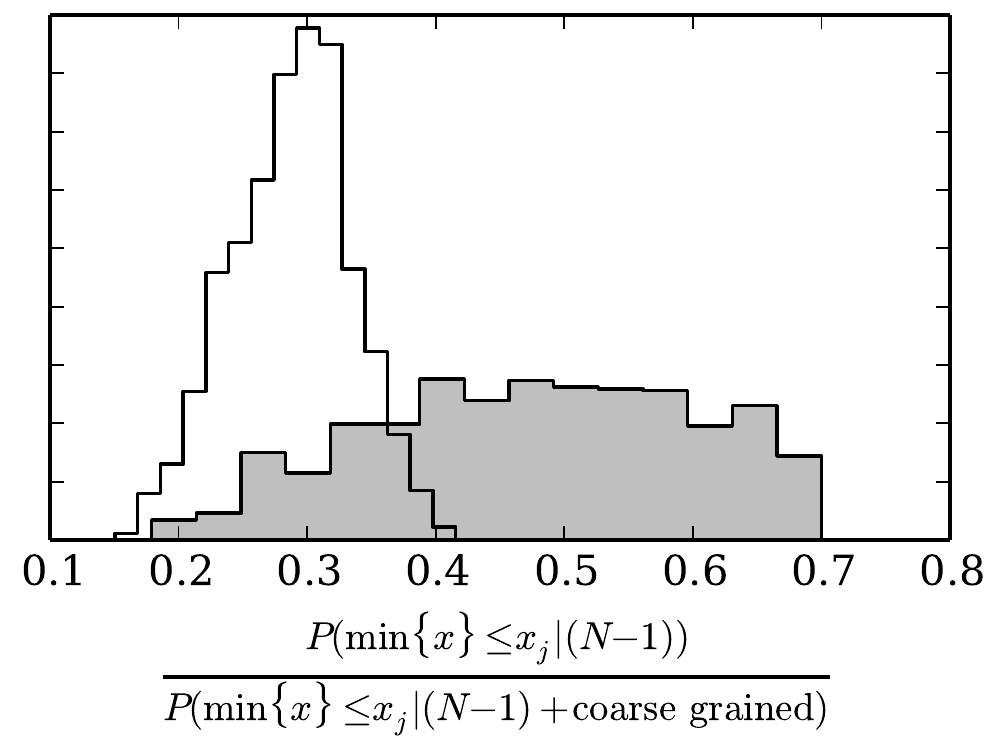}
        \end{center}
    \end{minipage}
    \caption{
        (\emph{left}) Distributions of probabilities that the smallest event in a catalog of \result{10} events would have $x \leq x_j$ ($p$-values), marginalized over hyperposteriors inferred with the other $N-1$ events when (\emph{shaded}) all events are drawn from the same distribution and (\emph{unshaded}) in the presence of a true outlier.
        Colors match Figs.~\ref{fig:toy model x hyperposteriors} and~\ref{fig:toy model q hyperposteriors}.
        (\emph{right}) Distributions of ratios of $p$-values from the different ($N-1$)-event hyperposteriors when the null hypothesis is (\emph{shaded}) true or (\emph{unshaded}) false.
        These differences are typically no larger than a factor of a few, although they are more pronounced when the null hypothesis is false.
        We also note that our $p$-values are only expected to be uniformly distributed when $\Lambda=\Lambda_0$ (see, e.g.,~\citet{10.1214/18-BA1124}).
        Because we marginalize over our uncertainty in $\Lambda$, the resulting statistic need not be uniformly distributed even when the null hypothesis is true.
    }
    \label{fig:toy model pvalues}
\end{figure*}

We investigate several astrophysical events from GWTC-2 in Sec.~\ref{sec:results}, but first consider the toy models in Sec.~\ref{sec:toy model excluding smallest x} and~\ref{sec:toy model excluding smallest q} in more detail.
In particular, we are concerned with changes in the probability of making \emph{type 1 errors} when the null hypothesis is true (incorrectly rejecting the null hypothesis, or a false positive).
At the same time, we are also interested in changes in the probability of making \emph{type 2 errors} when the null hypothesis is false (incorrectly failing to reject the null hypothesis, or a false negative).

Fig.~\ref{fig:toy model pvalues} shows the distribution of such $p$-values for different realizations of synthetic catalogs when we exclude events with the smallest $x$ (Sec.~\ref{sec:toy model excluding smallest x}).
We immediately note that the ($N-1$)-event inference, which neglects the coarse-graining correction, always predicts smaller $p$-values than the coarse-grained inference in both cases.
In this way, analysts could be tricked into claiming more significant tension than actually exists between the inferred model and the excluded event if they do not account for how the event was selected.
Similarly, they may be more likely to accept the null hypothesis when it is false based on coarse-grained inferences.
In either case, though, the $p$-values differ by only a factor of a few.
This may be why we reach the same astrophysical conclusions as \citet{O3aRatesAndPop} even though they did not include coarse-graining corrections; the size of the effect on $p$-values is nontrivial but still relatively modest.

%----------------------------------------------------------------------------------------

\section{Astrophysical Results}
\label{sec:results}

Using the coarse-grained inference described in Sec.~\ref{sec:methods} and our intuition from the toy models investigated in Sec.~\ref{sec:toy models}, we revisit several astrophysical events from GWTC-2.
We are specifically interested in evidence that individual events are incompatible with the main BBH population (the phenomenological distribution that describes the majority of detected BBH systems) inferred in~\citet{O3aRatesAndPop}. We consider GW190814 (Sec.~\ref{sec:GW190814}), GW190412 (Sec.~\ref{sec:GW190412}), and GW190521 (Sec.~\ref{sec:GW190521}) in turn.
Our analysis reweighs publicly available population hyperposteriors samples~\citep{O3aRatesAndPopDataRelease}; see Appendix~\ref{sec:reweighing} for more details.

In what follows, we focus on results with the \textsc{Powerlaw+Peak} mass model from~\citet{Talbot_2018, O3aRatesAndPop}.
Unless otherwise noted (i.e., GW190521 in Sec.~\ref{sec:GW190521}), astrophysical conclusions are unchanged when we assume different mass models.
Furthermore, we also consider fixed $\mathcal{S}$ in each case. % rather than marginalizing over the uncertainty in $\{\theta_{i\neq j}\}$.
Specific choices for $\mathcal{S}$ are listed in each section and are either motivated by the other $N-1$ events or theoretical expectations for astrophysical systems.

%------------------------

\subsection{GW190814 is an outlier in secondary mass}
\label{sec:GW190814}

We begin by considering GW190814~\citep{GW190814, GW190814DataRelease}, the BBH system with the smallest secondary mass and the most extreme mass ratio observed to date.
Indeed, GW190814's secondary is so small ($m_2\sim 2.6\,M_\odot$) that there has been significant discussion about whether it could have been a neutron star~\citep[see, e.g.,][]{Essick2020}, with the common consensus that the system is likely incompatible with a slowly-rotating neutron star.
For simplicity, we eschew the question of whether both components of GW190814 were actually black holes, and instead focus on whether their masses are compatible with the distribution inferred from the rest of the BBH events in GWTC-2.

\begin{figure*}
    \begin{minipage}{0.39\textwidth}
        \includegraphics[width=1.0\textwidth, clip=True, trim=0.0cm 1.40cm 0.00cm 0.00cm]{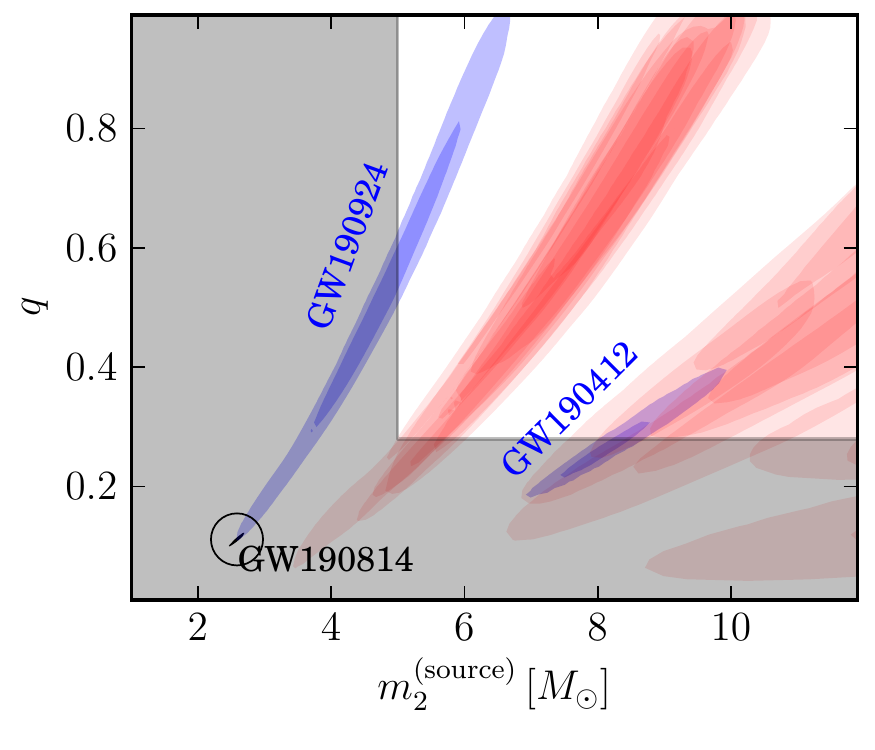} \\
        \includegraphics[width=1.0\textwidth, clip=True, trim=0.0cm 0.00cm 0.00cm 0.00cm]{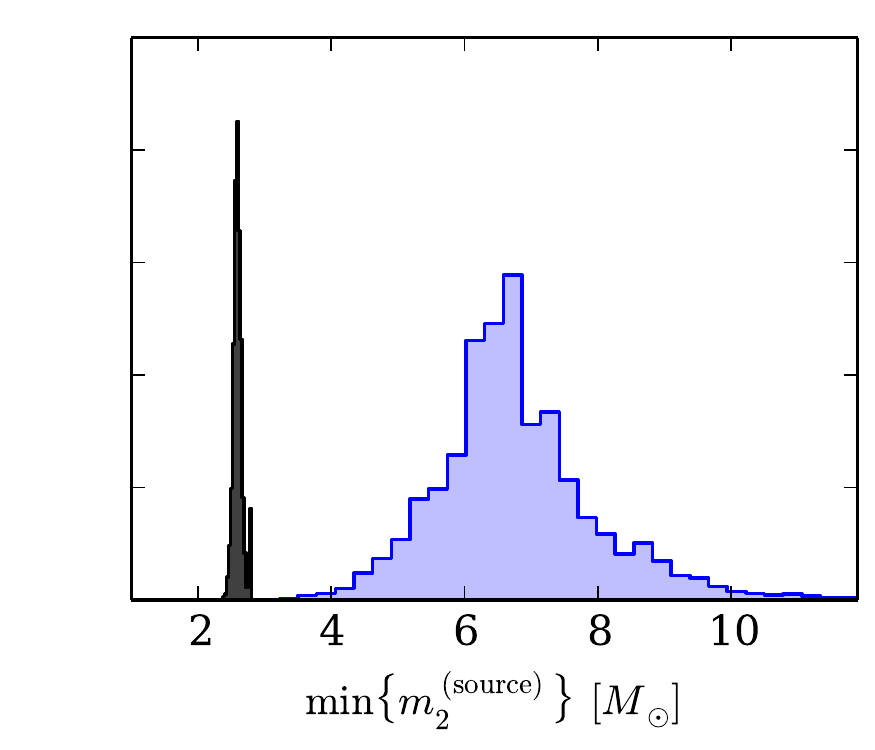} \\
    \end{minipage}
    \begin{minipage}{0.60\textwidth}
        \includegraphics[width=1.0\textwidth, clip=True, trim=1.0cm 0.5cm 0.5cm 0.5cm]{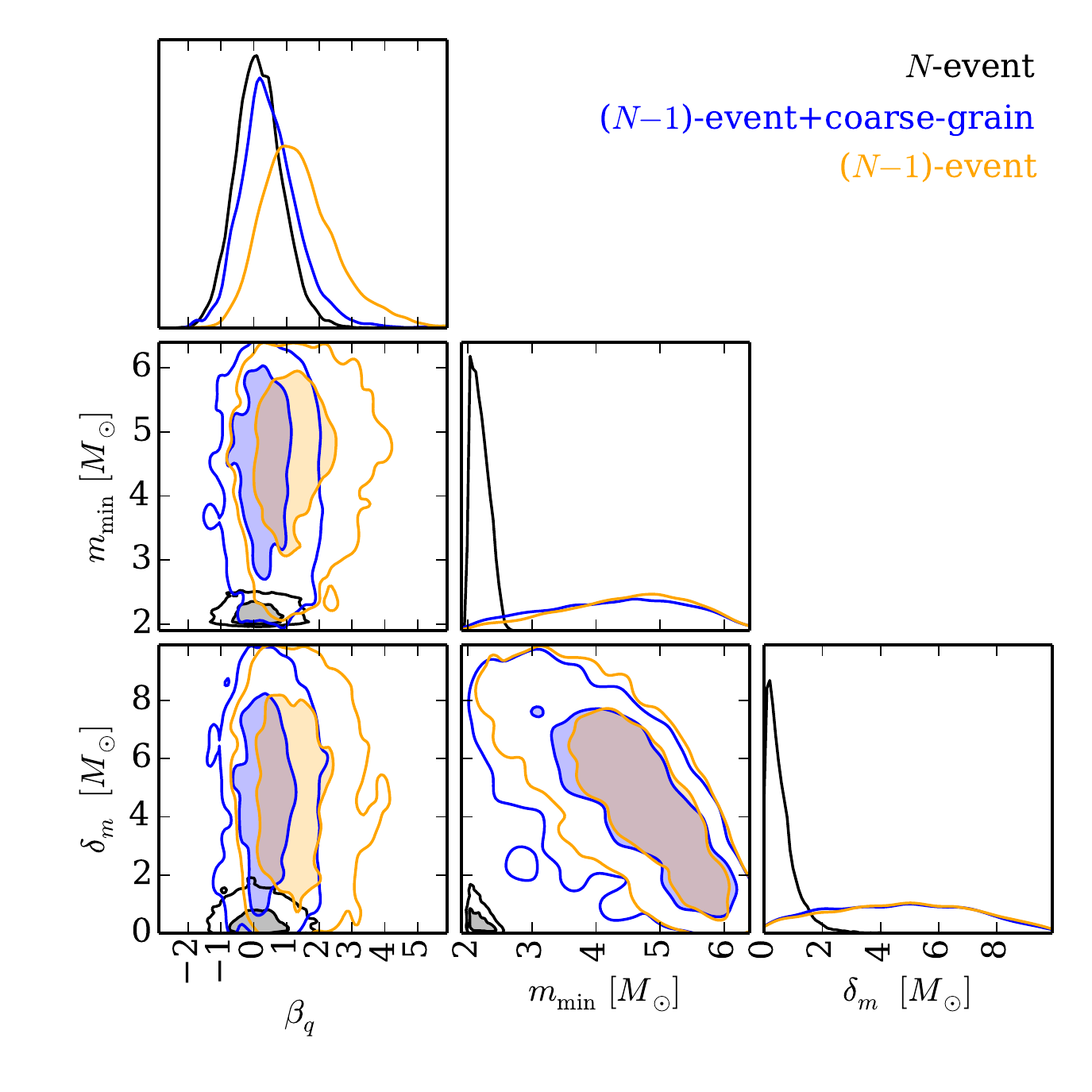}
    \end{minipage}
    \caption{
        (\emph{top left}) Depiction of $\mathcal{S}_\mathrm{GW190814}$ (\emph{shaded region}, Eq.~\ref{eq:S_GW190814}), chosen to be the region $(m_2 \leq \FourteenMassCut)$ \emph{or} $(q \leq \FourteenQCut)$ based on the approximate median $m_2$ and $q$ of GW190924 and GW190412, respectively.
        Colored shaded regions show 50\% and 90\% highest-probability-density credible regions from each event assuming flat priors in component masses and uniform priors in comoving volume for (\emph{black} ellipse, circled in lower left corner) GW190814, (\emph{blue}) GW190924 and GW190412, as well as (\emph{red}) all other BBH systems considered in~\citet{O3aRatesAndPop}.
        (\emph{bottom left}) Distribution of (\emph{black}) GW190814's $m_2$ and (\emph{blue}) the smallest expected $m_2$ in catalogs of \result{45} events based on the ($N-1$)-event coarse-grained hyperposterior for the \textsc{PowerLaw+Peak} mass model.
        (\emph{right}) Hyperposteriors inferred with different amounts of information about GW190814. %~\citep{GW190814}.
        Contours in the joint distributions denote 50\% and 90\% credible regions.
    }
    \label{fig:GW190814}
\end{figure*}

We adopt the mass models explored in~\citep{O3aRatesAndPop}, all of which include a cut-off at low masses (albeit with variable degrees of sharpness) in much the same way as our toy model (Sec.~\ref{sec:toy models}).
In some sense, then, the results in Fig.~\ref{fig:GW190814} are directly comparable to Fig.~\ref{fig:toy model x hyperposteriors}.

We define $\mathcal{S}$ for our analysis of GW190814 as follows
\begin{equation}\label{eq:S_GW190814}
    \mathcal{S}_\mathrm{GW190814} : (m_2 \leq \FourteenMassCut)\ \mathrm{OR}\ (q \leq \FourteenQCut) .
\end{equation}
The $m_2\leq\FourteenMassCut$ boundary is chosen to match the median posterior estimate of GW190924, the event with the second smallest secondary mass after GW190814.
The $q\leq\FourteenQCut$ boundary is chosen to match the median posterior estimate of GW190412, the event with the second smallest mass ratio after GW190814.
Both GW190924 and GW190814 are highlighted in blue in Fig.~\ref{fig:GW190814}.
This defines an ``L-shaped'' region in the ($m_2$, $q$) plane spanning the lowest values for both dimensions (see Fig.~\ref{fig:GW190814}).
We again note that this choice of $\mathcal{S}$ is not unique, and one could instead choose to define $\mathcal{S}$ with bounds on only $m_2$ or only $q$.
Defining $\mathcal{S}_\mathrm{GW190814}$ in terms of only $m_2$ or only $q$ does not affect our conclusions, and defining $\mathcal{S}_\mathrm{GW190814}$ in this way allows us to be as agnostic as possible about GW190814 in some sense.
Importantly, we note that, with this definition of $\mathcal{S}_\mathrm{GW190814}$, population models that do not have support for masses below $m_2 \leq \FourteenMassCut$ are still allowed as long as they support $q \leq \FourteenQCut$, and \textit{vice versa}.

Fig.~\ref{fig:GW190814} summarizes our conclusions.
Specifically, we see that the ($N-1$)-event coarse-grained hyperposterior resembles the full $N$-event hyperposterior for $\beta_q$, the hyperparameter that controls the extent of the population model for $q$, while it more closely resembles the ($N-1$)-event hyperposterior that neglects the coarse-grained correction for both $m_\mathrm{min}$ and $\delta_m$, which control the minimum mass allowed in the population.
We note that $m_\mathrm{min}$ sets an absolute lower bound for the allowed masses within a population, and therefore we would expect a reasonable amount of probability that $m_\mathrm{min} \leq 3\,M_\odot$ in the ($N-1$)-event analyses if GW190814 was consistent with the population inferred from the other events.
While $m_\mathrm{min} \leq 3\,M_\odot$ is not excluded by the ($N-1$)-event analysis, it is not particularly favored, either.
Our intuition from Sec.~\ref{sec:toy models}, then, suggests that \emph{GW190814's small $q$ is consistent with the rest of the events} (population models already contain plenty of support for small $q$), but \emph{GW190814 is inconsistent with the rest of the events because of its small $m_2$}.

We further quantify this by estimating a $p$-value that the smallest event out of an $N$-event catalog would have $m_2$ less than or equal to any event in GWTC-2, given the ($N-1$)-event coarse-grained hyperposterior.
We find $P\, \FourteenPowerlawPeakPvalCoarseGrained$ at 90\% confidence.\footnote{We can only bound the $p$-value from above as we did not find a single instance where synthetic catalogs generated $\min\{m_2\}$ smaller than GW190814's secondary after $\sim 5000$ trials. Similarly, we find $P\, \FourteenPowerlawPeakPvalNminusOne$ with the ($N-1$)-event analysis that neglects coarse graining.}
As such, we reject the null hypothesis that GW190814 was drawn from the same population as the other $N-1$ events.
\citet{O3aRatesAndPop} reach the same conclusion, and, following their example, we exclude GW190814 from the catalog as we explore whether other individual events are inconsistent with the remaining detections.

%------------------------

\subsection{GW190412 is not an outlier}
\label{sec:GW190412}

\begin{figure}
    \includegraphics[width=1.0\columnwidth, clip=True, trim=0.0cm 0.0cm 0.0cm 0.0cm]{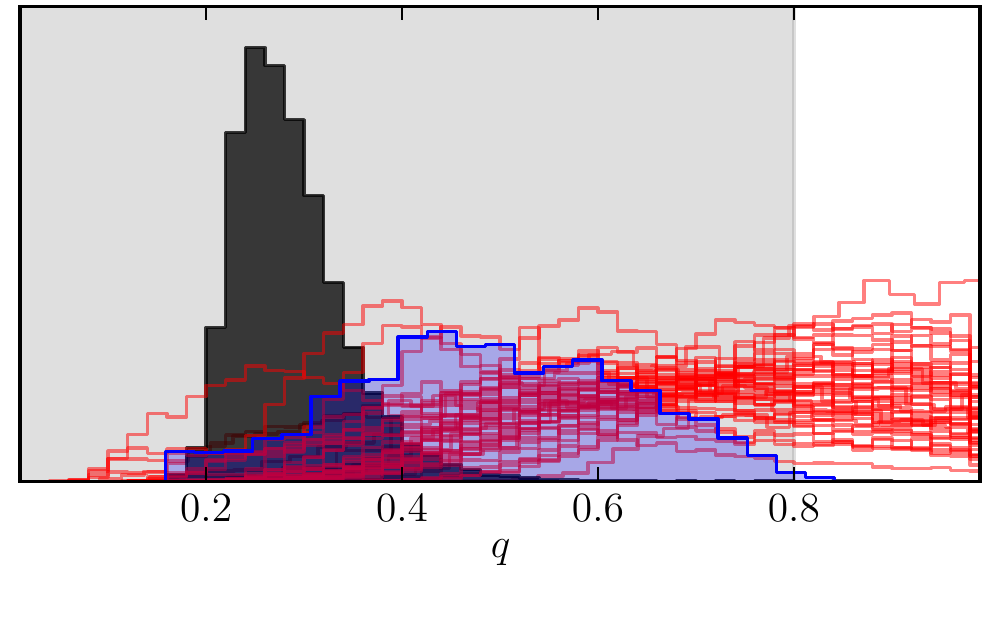}
    \includegraphics[width=1.0\columnwidth, clip=True, trim=0.0cm 0.0cm 0.0cm 0.0cm]{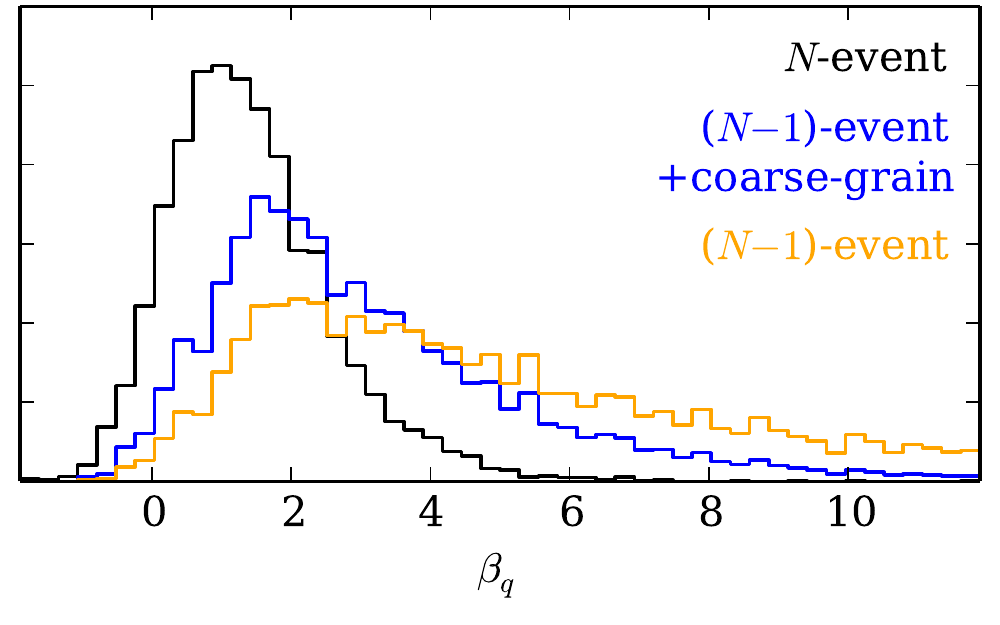}
    \caption{
        (\emph{top}) Depiction of (\emph{shaded grey}) $\mathcal{S}_\mathrm{GW190412}$ (Eq.~\ref{eq:S_GW190412}) with (\emph{shaded black}) GW190412 and (\emph{red}) all other events considered in~\citet{O3aRatesAndPop} (except GW190814) with flat priors on component masses as well as (\emph{blue}) the prediction for the smallest $q$ out of \result{44} events based on the ($N-1$)-event coarse-grained hyperposterior with the \textsc{PowerLaw+Peak} mass model.
        (\emph{bottom}) Hyperposteriors inferred with different amounts of information about GW190412.
    }
    \label{fig:GW190412}
\end{figure}

We next investigate GW190412~\citep{GW190412, GW190412DataRelease}, which, with $q \sim 0.28$, is the only event in GWTC-2 besides GW190814 that is inconsistent with $q = 1$.
As a reminder, we have already removed GW190814 from the set of events against which we compare GW190412.
In this case, we define
\begin{equation}\label{eq:S_GW190412}
    \mathcal{S}_\mathrm{GW190412} : q \leq \TwelveQCut .
\end{equation}
as a conservative boundary for systems that have asymmetric masses.
\citet{Fishbach2020Mar} estimate that \externalresult{90\% (99\%) of detected BBHs will have $q \geq $ 0.73 (0.51) based on population models of the 10 BBH events in GWTC-1~\citep{GWTC-1}}.
Our choice for $\mathcal{S}_\mathrm{GW190412}$ is even more conservative: we give the coarse-grained inference very little information about GW190412 itself, and will therefore most easily identify it as an outlier.
Fig.~\ref{fig:GW190412} demonstrates the results.

Again, we are primarily interested in the low-mass (and low mass-ratio) behavior of the model and focus on the inferred values of $\beta_q$, $m_\mathrm{min}$, and $\delta_m$.
In this case, we find that all inferences agree remarkably well for $m_\mathrm{min}$ and $\delta_m$, which is unsurprising since GW190412's masses are not  particularly extreme when considered individually.
However, we observe better agreement between the $N$-event and ($N-1$)-event coarse-grained hyperposteriors for $\beta_q$ than between either and the ($N-1$)-event inference that neglects coarse-graining.
This is analogous to Fig.~\ref{fig:toy model q hyperposteriors} when the null hypothesis is true; excluding the event with smallest $q$ shifts the inferred hyperposterior towards values that favor equal-mass systems (larger $\beta_q$).

The coarse-grained inference produces a $p$-value of $P = \TwelvePowerlawPeakPvalCoarseGrained$\footnote{The ($N-1$)-event analysis without coarse graining yields $P = \TwelvePowerlawPeakPvalNminusOne$.} for the smallest observed $q$ to be as small or smaller than that of any BBH event in GWTC-2 (excluding GW190814).
We therefore conclude that \emph{GW190412 is consistent with the population inferred from rest of the BBH events within GWTC-2} (except GW190814).
It is simply the event with the most extreme $q$ from that population, in agreement with both~\citet{GW190412} and ~\citet{O3aRatesAndPop}.

%------------------------

\subsection{GW190521 is not an outlier}
\label{sec:GW190521}

\begin{figure}
    \includegraphics[width=1.0\columnwidth, clip=True, trim=0.0cm 0.00cm 0.0cm 0.0cm]{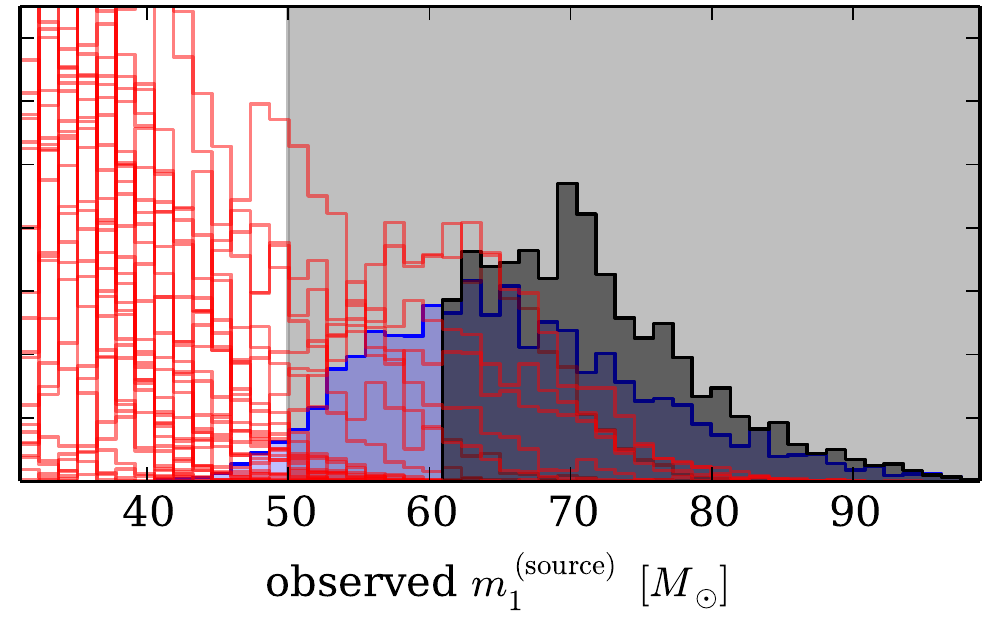}
    \includegraphics[width=1.0\columnwidth, clip=True, trim=0.0cm 0.00cm 0.0cm 0.0cm]{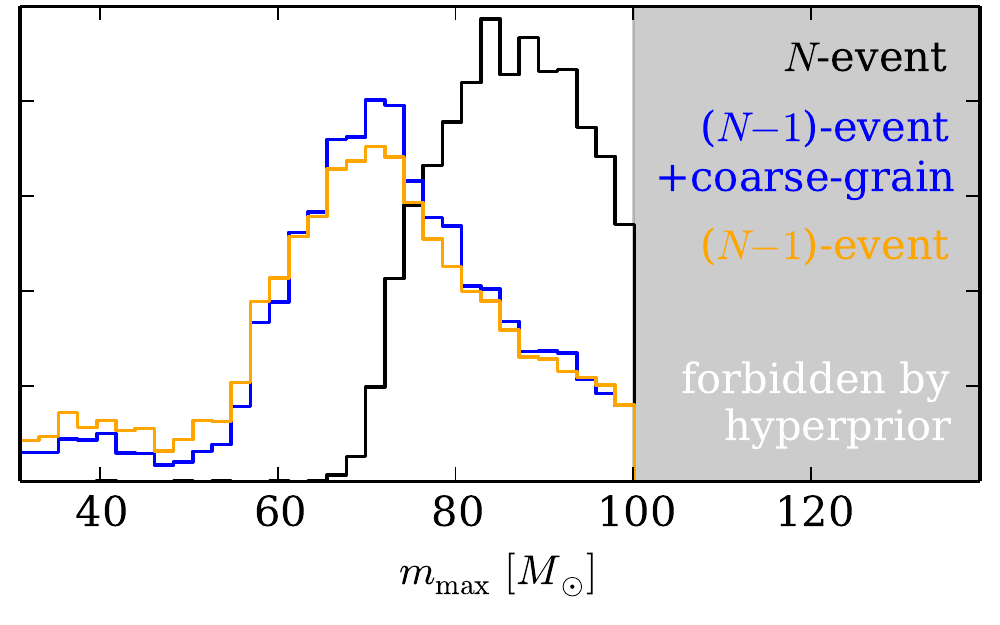}
    \caption{
        (\emph{top}) Depiction of (\emph{shaded grey}) $\mathcal{S}_\mathrm{GW190521}$ (Eq.~\ref{eq:S_GW190521}) with (\emph{black shaded}) GW190521 and  (\emph{red}) all other events considered in~\citep{O3aRatesAndPop} informed by the \textsc{PowerLaw+Peak} mass model as inferred under the ($N-1$)-event coarse-grained hyperposterior as well as the distribution of the largest predicted $m_1$ out of \result{44} events based on the ($N-1$)-event coarse-grained hyperposterior.
        The hard boundary in GW190521's distribution at $m_1 \sim 60\,M_\odot$ is due to our labeling convention ($m_1 \geq m_2$) and approximately corresponds to equal-mass systems.
        (\emph{bottom}) Hyperposteriors inferred with different amounts of information about GW190521.
        Note that $m_\mathrm{max}$ extends below $50\,M_\odot$ because it only limits the power law part of \textsc{Powerlaw+Peak}; the peak still has support at higher masses.
    }
    \label{fig:GW190521}
\end{figure}

Finally, we consider GW190521~\citep{GW190521, GW190521Properties, GW190521DataRelease}.
This event is remarkable for its large component masses, although we note that it does not unambiguously have the largest component masses of any system in GWTC-2; see Fig.~\ref{fig:GW190521}.
In this case, we cannot easily define $\mathcal{S}_\mathrm{GW190521}$ in terms of the other events in the catalog while guaranteeing that Eq.~\ref{eq:minimum S} is satisfied.
However, we note that GW190521 is of particular interest because its component masses nominally fall within the PISN mass gap.\footnote{See, e.g., \citet{Fishbach2020Nov} for alternative interpretations with component masses that straddle the mass gap. 
}
In this case, it is natural to define $\mathcal{S}_\mathrm{GW190521}$ in terms of an approximate boundary defining the PISN mass gap.
We take
\begin{equation}\label{eq:S_GW190521}
    \mathcal{S}_\mathrm{GW190521} : m_1 \geq \TwentyOneMassCut
\end{equation}
as a reasonable approximation, although others have considered values as large as $65\,M_\odot$~\citep{GW190521, GW190521Properties, OBrien2021}.
Again, there is nontrivial overlap between this choice for $\mathcal{S}_\mathrm{GW190521}$ and the parameters inferred for several other events in GWTC-2, but this does not affect our inference.

Contrary to GW190814 and GW190412, in which higher-order modes were observed and their asymmetric mass ratios were relatively well measured, GW190521's component mass posterior is quite broad.
As such, simultaneous population inference can have a significant impact on the system's inferred properties.
For this reason, Fig.~\ref{fig:GW190521} shows the posteriors of primary masses under the \textsc{PowerLaw+Peak} mass model as inferred with our coarse-grained analysis.
Nonetheless, several analyses have shown that it is likely that at least one component of GW190521 had a mass $\geq 65\,M_\odot$~\citep{GW190521, GW190521Properties, OBrien2021}.

Some authors have taken this to mean that GW190521 is incompatible with the population of stellar-mass black holes, but to be clear they often mean the inferred phenomenological fit from other events with an additional cut-off (not included in the fit).
We are concerned with the simpler question of whether GW190521 is inconsistent with the phenomenological population model inferred from the other BBH events without imposing additional sharp cut-offs.
We are not concerned with whether the current phenomenological models are compatible with PISN predictions, but instead ask whether the models are sufficient to describe the overall mass distribution and whether GW190521 is consistent with models inferred from the other events.

While our coarse-grained hyperposterior at times more closely resembles the ($N-1$)-event hyperposterior than the $N$-event hyperposterior (bottom panel of Fig.~\ref{fig:GW190521}), we nonetheless obtain a $p$-value of $P=\TwentyOnePowerlawPeakPvalCoarseGrained$ that synthetic catalogs would contain a more massive event than has been observed so far.
In fact, if we neglect the coarse graining correction, we still obtain $P=\TwentyOnePowerlawPeakPvalNminusOne$ under the \textsc{PowerLaw+Peak} model.
As such, and in agreement with~\citet{O3aRatesAndPop}, we conclude that GW190521 is consistent with the overall mass distribution inferred from the rest of the BBH population.

One might expect other mass models, like~\citet{O3aRatesAndPop}'s \textsc{Truncated} model with a sharp cut-off at high masses, to be less consistent with GW190521.
To investigate this, we repeat our coarse-grained analysis and find $P=\TwentyOneTruncatedPvalCoarseGrained$ under the \textsc{Truncated} model.
Indeed, even the ($N-1$)-event \textsc{Truncated} hyperposterior that neglects coarse-graining only yields $P=\TwentyOneTruncatedPvalNminusOne$.
As such, we may conclude that, while GW190521 is certainly not expected to be common, it also is not unambiguously inconsistent with any of the phenomenological models considered in~\citet{O3aRatesAndPop}, even the simplest \textsc{Truncated} distribution.\footnote{While GW190521 may not be inconsistent with the mass models considered in~\citet{O3aRatesAndPop}, they point out that the simple \textsc{Truncated} model is more broadly a bad fit to the data. It overpredicts the number of massive systems that should have been observed (see also~\cite{Fishbach2021}). However, even minor modifications like the \textsc{PowerLaw+Peak} and \textsc{Broken PowerLaw} distributions appear to remove such tensions.}

That being said, the number of BBH systems detected in GWTC-2 more than quadrupled compared to GWTC-1, and both \citet{O3aRatesAndPop} and \cite{Fishbach2020Nov} point out that GW190521 does seem to be inconsistent with the truncated mass distributions inferred based on only the 10 events in GWTC-1~\citep{2019ApJ...882L..24A}.
While additional GW detections have continued to surprise, we are reminded that it is important to consider new events in the context of the full catalog before drawing conclusions based on individual (apparently) exceptional events and a subset of previously observed systems.

%----------------------------------------------------------------------------------------

\section{Discussion}
\label{sec:discussion}

Within any set of observed events drawn from an unknown population, one is often interested in determining whether new events are consistent with the population inferred from the existing set.
This often involves careful examination of particular events because they are extremal in some way.
It is remarkable that GW astronomy has already advanced to the point where such matters are of practical importance in only a half-dozen years since the first detection~\citep{GW150914}.
Nonetheless, we show that current approaches to answer exactly this question, which have become commonplace within the GW community, can introduce biases as they do not account for the manner in which extremal events were identified for further study.

Our method allows analysts to explicitly account for how they selected extremal events within leave-one-out analyses, representing excluded events with coarse-grained likelihoods, and clearly identifying the need to select the size and placement of the coarse grains.
While we note that the exact choice of how big to make those grains is to some degree arbitrary, just as the definition of a null hypothesis is to some degree arbitrary, we propose algorithmic ways to choose the most generous grains possible.
We further observe that the resulting coarse-grained analysis almost always has nearly correct coverage within several toy models.

Finally, we note that the biases introduced by excluding extremal events without accounting for how they were selected can be particularly severe when there are sharp features in the underlying population model (e.g., mass gaps).
Therefore, one must take care when analyzing population models with sharp cut-offs and attempting to assess the significance of outliers after excluding extremal events.
However, even in these severe cases, we find that $p$-values estimated from ($N-1$)-event hyperposteriors are typically biased by at most a factor of a few.

Our conclusions based on our coarse-grained analysis agree with those presented in~\citet{O3aRatesAndPop}, even though they did not account for the coarse-grained correction and their leave-one-out analyses may have been biased.
We find that \emph{GW190814 is an outlier} because its secondary mass is too small to be consistent with the other events.
\emph{GW190412 is not an outlier}, as its small mass ratio is simply the most extreme example from the tail of the main population.
We find that \emph{GW190521 is not an outlier} under the preferred mass models explored in~\citet{O3aRatesAndPop}, and is in only moderate tension with even the simplest truncated mass models.

We again note that any population analysis will eventually face the challenge of determining whether particular events are consistent with the population inferred from the rest of the events.
This problem is not unique to GW astronomy.
However, as catalogs continue to rapidly grow in size, this question has become increasingly relevant.
We note that another large set of events is expected with the release of the second half of the LVK collaborations' third observing run (O3b).
Indeed, given the breadth of physical phenomena that GW observations can probe, it is of the utmost importance to fully characterize outlier tests.
We emphasize that many of the most interesting questions in GW astronomy are specifically focused on outliers, including extreme mass, mass ratio, and spin events, and the presence of events within the putative NS-BH and PISN mass gaps.
Our analysis provides a controlled way to account for event selection when examining outliers with nearly trivial additional computational cost.
This will enable the robust identification of novel subpopulations without fear of biasing analyses towards artificially inflated significance estimates for potential outliers.

%----------------------------------------------------------------------------------------

\acknowledgements

The authors thank Will Farr, Tom Callister, and Katerina Chatziioannou for several helpful discussions.
Research at Perimeter Institute is supported in part by the Government of Canada through the Department of Innovation, Science and Economic Development Canada and by the Province of Ontario through the Ministry of Colleges and Universities.
S.G. and E.T. are supported through Australian Research Council (ARC) Centre of Excellence CE170100004.
A.F. is supported by the NSF Research Traineeship program under grant DGE-1735359.
M.F. is supported by NASA through NASA Hubble Fellowship grant HST-HF2-51455.001-A awarded by the Space Telescope Science Institute.
D.E.H is supported by NSF grants PHY-2006645, PHY-2011997, and PHY-2110507, as well as by the Kavli Institute for Cosmological Physics through an endowment from the Kavli Foundation and its founder Fred Kavli.
D.E.H also gratefully acknowledges the Marion and Stuart Rice Award.
This material is based upon work supported by NSF LIGO Laboratory which is a major facility fully funded by the National Science Foundation.
The authors are grateful for computational resources provided by the LIGO Laboratory and supported by National Science Foundation Grants PHY-0757058 and PHY-0823459.

%----------------------------------------------------------------------------------------

\bibliography{references}

%----------------------------------------------------------------------------------------

\appendix

%--------------------------------------

\section{Reweighing Existing Hyperposteriors}
\label{sec:reweighing}

We note that Eq.~\ref{eq:coarse-grained post} could be computationally expensive if we allow $\mathcal{S}$ to depend on the parameters of the other events.
However, if $\mathcal{S}$ does not depend on the parameters of the other events, the marginalization becomes trivial.
There are also significant redundancies between Eq.~\ref{eq:full post} and Eq.~\ref{eq:coarse-grained post}, which reduce the computational cost of reweighing existing samples and make implementing the coarse-grained likelihood a simple extension of existing likelihoods.

\begin{figure}[hb]
    \centering
    \includegraphics[width=0.54\textwidth, clip=True, trim=1.0cm 0.5cm 0.5cm 0.5cm]{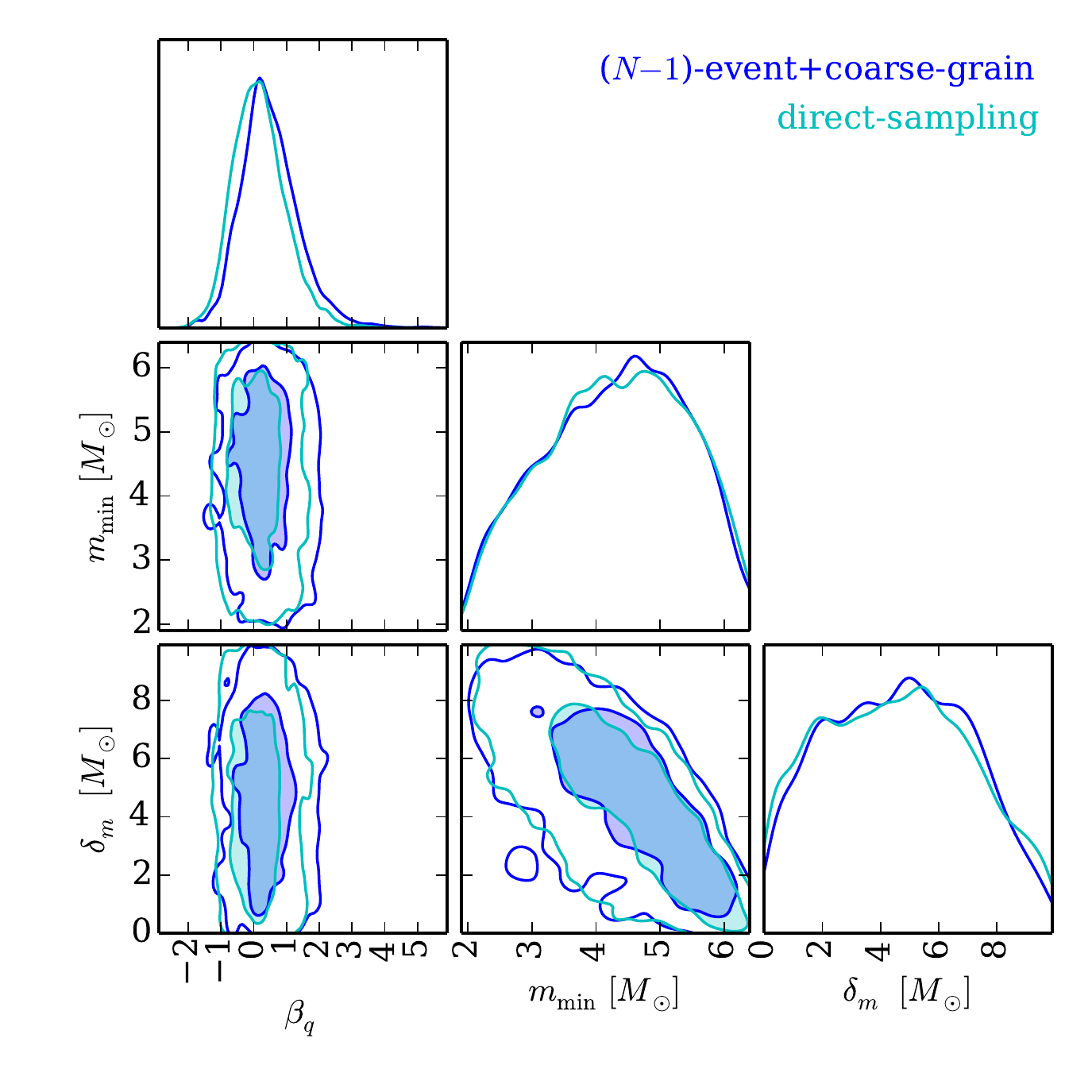}
    \caption{
        Comparison between reweighed samples and direct sampling for GW190814's ($N-1$)+coarse-grained hyperposterior (Sec.~\ref{sec:GW190814}).
        Contours in the joint distributions denote 50\% and 90\% credible regions.
        Reweighed samples were obtained by applying Eq.~\ref{eq:reweigh N-1 to coarse-grained} to public hyperposterior samples from an ($N-1$)-event analysis that excluded GW190814~\cite{O3aRatesAndPopDataRelease}.
    }
    \label{fig:GW190814_direct}
\end{figure}

There are two corrections to the likelihood that may be of interest.
When we have already analyzed the full set of $N$ events and want to conduct a coarse-grained analysis \textit{post hoc}, we note that
\begin{equation}\label{eq:reweight N to coarse-grained}
    \frac{p(\Lambda|\{D_{i\neq j}\}, \rho(D_j)\geq\rho_\mathrm{thr}; \theta_j\in\mathcal{S})}{p(\Lambda|\{D_{i}\})} \propto
    \frac{
    \int d\theta\, p(\theta|\Lambda) P(\mathrm{det}|\theta) \Theta(\theta\in\mathcal{S}(\{\theta_{i\neq j}\}))
    }{
    \int d\theta_j\, p(D_j|\theta_j)p(\theta_j|\Lambda)
    } = \frac{P(\theta \in \mathcal{S}, \mathrm{det}|\Lambda)}{p(D_j|\Lambda)}.
\end{equation}
When we instead have hyperposterior samples from an ($N-1$)-event analysis that omitted the potential outlier, we can write
\begin{equation}\label{eq:reweigh N-1 to coarse-grained}
    \frac{p(\Lambda|\{D_{i\neq j}\}, \rho(D_j)\geq\rho_\mathrm{thr}; \theta_j\in\mathcal{S})}{p(\Lambda|\{D_{i\neq j}\})} \propto P(\theta\in\mathcal{S}|\mathrm{det};\Lambda).
\end{equation}
Weighing existing hyperposterior samples by either Eq.~\ref{eq:reweight N to coarse-grained} or~\ref{eq:reweigh N-1 to coarse-grained}, as appropriate, allows us to estimate the coarse-grained hyperposterior and quickly perform consistency tests of our null hypothesis.
Such reweighing procedures are equivalent to directly sampling from Eq.~\ref{eq:coarse-grained post} as long as enough effective samples remain to provide reliable estimates of the hyperposterior.
As a demonstration, we repeat the analysis of Sec.~\ref{sec:GW190814} by directly sampling from Eq.~\ref{eq:coarse-grained post}, obtaining equivalent results to the reweighed hyperposterior samples (see Fig.~\ref{fig:GW190814_direct}).

%----------------------------------------------------------------------------------------
\end{document}